\documentstyle[amssymb,psfig]{mn}

\newcommand{\tdisref}{\mbox{$t_4^{\rm dis}$}}

\title[Star Cluster Formation and Evolution in Nearby Starburst
Galaxies: II.]{Star Cluster Formation and Evolution in Nearby Starburst
Galaxies: II. Initial Conditions}

\author[R. de Grijs et al.]{R. de Grijs$^1$\thanks{E-mail: 
grijs@ast.cam.ac.uk}\thanks{Present address: Department of Physics \&
Astronomy, University of Sheffield, Hicks Building, Hounsfield Road,
Sheffield S3 7RH}, P. Anders,$^2$ N. Bastian,$^3$ R. Lynds,$^4$ 
H.J.G.L.M. Lamers,$^3$ and
\newauthor E.J. O'Neil, Jr.$^4$
\\ 
$^1$ Institute of Astronomy, University of Cambridge, Madingley Road,
Cambridge CB3 0HA\\
$^2$ Universit\"atssternwarte, University of G\"ottingen,
Geismarlandstr. 11, 37083 G\"ottingen, Germany\\
$^3$ Astronomical Institute, Utrecht University, Princetonplein 5,
3584 CC Utrecht, The Netherlands \\
$^4$ Kitt Peak National Observatory, National Optical Astronomy
Observatories, Box 26732, Tucson, AZ 85726, USA
}

\date{Received date; accepted date}
\pubyear{2003}

\begin{document}
\maketitle

\begin{abstract}
We use the ages, masses and metallicities of the rich young star cluster
systems in the nearby starburst galaxies NGC 3310 and NGC 6745 to derive
their cluster formation histories and subsequent evolution.  We further
expand our analysis of the systematic uncertainties involved in the use
of broad-band observations to derive these parameters (Paper I) by
examining the effects of {\it a priori} assumptions on the individual
cluster metallicities. 
The age (and metallicity) distributions of both the clusters in the
circumnuclear ring in NGC 3310 and of those outside the ring are
statistically indistinguishable, but there is a clear and significant
excess of higher-mass clusters {\it in} the ring compared to the
non-ring cluster sample; it is likely that the physical conditions in
the starburst ring may be conducive for the formation of higher-mass
star clusters, on average, than in the relatively more quiescent
environment of the main galactic disc. 
For the NGC 6745 cluster system we derive a median age of $\sim 10$
Myr.  NGC 6745 contains a significant population of high-mass ``super
star clusters'', with masses in the range $6.5 \lesssim \log( M_{\rm
cl}/M_\odot ) \lesssim 8.0$.  This detection supports the scenario that
such objects form preferentially in the extreme environments of
interacting galaxies. 
The age of the cluster populations in both NGC 3310 and NGC 6745 is
significantly lower than their respective characteristic cluster
disruption time-scales, respectively $\log(t_4^{\rm dis}/{\rm yr}) =
8.05$ and 7.75, for $10^4 M_\odot$ clusters.  This allows us to obtain
an independent estimate of the {\it initial} cluster mass function
slope, $\alpha = 2.04(\pm 0.23)^{+0.13}_{-0.43}$ for NGC 3310, and $1.96
(\pm 0.15)\pm 0.19$ for NGC 6745, respectively, for masses $M_{\rm cl}
\gtrsim 10^5 M_\odot$ and $M_{\rm cl} \gtrsim 4 \times 10^5 M_\odot$. 
These mass function slopes are consistent with those of other young star
cluster systems in interacting and starburst galaxies. 
\end{abstract}


\section{Introduction}
\label{intro.sec}

\subsection{Star cluster formation in intense starbursts}

The production of luminous, compact star clusters seems to be a hallmark
of intense star formation. Such young clusters have been identified in
intense starburst regions in several dozen galaxies, often involved in
interactions (e.g., Holtzman et al. 1992, Whitmore et al. 1993,
O'Connell et al. 1994, Conti et al. 1996, Watson et al. 1996, Carlson
et al. 1998, de Grijs et al. 2001, 2003d). Their sizes, luminosities,
and -- in several cases -- spectroscopic mass estimates are entirely
consistent with what is expected for young Galaxy-type globular clusters
(GCs; Meurer 1995, van den Bergh 1995, Ho \& Filippenko 1996a,b,
Schweizer \& Seitzer 1998, de Grijs et al. 2001, 2003a). 

It is possible, even likely, that a large fraction of the star formation
in starbursts takes place in the form of such concentrated clusters
(see, e.g., de Grijs et al.  2001, 2003d for a discussion).  Young
compact star clusters are therefore important because of what they can
tell us about GC formation and evolution (e.g., destruction mechanisms
and efficiencies).  They are also important as probes of the history of
star formation, chemical evolution, initial mass function (IMF), and
other physical characteristics in starbursts.  This is so because each
cluster approximates a coeval, single-metallicity, simple stellar
population (SSP).  Such systems are the simplest to model, after
individual stars themselves, and their ages and metallicities and -- in
some cases -- IMFs can be estimated from their integrated spectra. 

In de Grijs et al. (2003c; hereafter Paper I) we developed a reliable
method to determine cluster ages, masses, metallicities and extinction
values robustly and simultaneously based on imaging observations in a
minimum of four broad-band passbands covering the entire optical
wavelength range from the {\it U} to the {\it I} band or their
equivalents in non-standard passband systems. We tested our method
using the $\sim 150$ clusters in the centre of the nearby starburst
galaxy NGC 3310, and confirmed the previously suggested scenario that
NGC 3310 underwent a (possibly extended) global burst of cluster
formation $\sim 3 \times 10^7$ yr ago, likely triggered by the last
tidal interaction or merger with a low-metallicity, gas-rich dwarf
galaxy (see Paper I). 

However, in determining the cluster formation history from the age
distribution of magnitude-limited cluster samples, cluster disruption
must be taken into account.  This is because the observed age
distribution is that of the surviving clusters only.  In this paper, we
will therefore embark on a further analysis of the formation history of
the NGC 3310 cluster system and its subsequent evolution.  We will also
apply our method to the star clusters and star-forming regions in the
interacting galaxy NGC 6745 (the ``Bird's Head Galaxy''), and interpret
its cluster formation history in the contexts of its recent tidal
encounter and of cluster disruption. 

\subsection{An empirical description of cluster disruption}

As shown by Boutloukos \& Lamers (2003, hereafter BL03), and applied to
the fossil starburst cluster sample in M82 B by de Grijs et al. 
(2003b), with only a few well-justified assumptions, the mass and age
distributions of a magnitude-limited sample of star clusters in a given
galaxy can be predicted both accurately and robustly, despite the
complex physical processes underlying the assumptions (for a full
discussion see BL03). If all of the following conditions are met then
it can be shown easily that the age distribution of the observed cluster
population will obey approximate power-law behaviours. 

\begin{enumerate}
\item The cluster formation rate, ${{\rm d}N(M_{\rm cl}) \over {\rm d}t}
= S \times M_{\rm cl}^{-\alpha}$, is constant;

\item The slope $\alpha$ of the cluster IMF is constant among cluster
systems, with $N(M_{\rm cl}) {\rm d}M \propto M_{\rm cl}^{-\alpha} {\rm
d}M$;

\item Stellar evolution causes SSPs to fade as $F_{\lambda} \sim
t^{-\zeta}$, as predicted by theoretical cluster evolution models; and

\item The cluster disruption time depends on their initial mass as
$t_{\rm dis} = \tdisref \times (M_{\rm cl} / 10^4 M_\odot)^\gamma$,
where $\tdisref$ is the disruption time-scale of a cluster with initial
mass $M_{\rm cl} = 10^4 M_\odot$. 

\end{enumerate} 
It is well-established, however, that the disruption time-scale does not
only depend on mass, but also on the initial cluster density and
internal velocity dispersion (e.g., Spitzer 1957, Chernoff \& Weinberg
1990, de la Fuente Marcos 1997, Portegies Zwart et al. 2001). 
Following the approach adopted in BL03 and de Grijs et al. (2003b), we
point out that if clusters are approximately in pressure equilibrium
with their environment, we can expect the density of all clusters in a
limited volume of a galaxy to be roughly similar, so that their
disruption time-scale will predominantly depend on their (initial) mass
(with the exception of clusters on highly-eccentric orbits). In the
opposite case that the initial cluster density $\rho$ depends on their
mass $M_{\rm cl}$ in a power-law fashion, e.g., $\rho \sim M_{\rm cl}^x$
with $x$ being the (arbitrary) power-law exponent, the disruption
time-scale will also depend on mass if $t_{\rm dis} \sim M_{\rm cl}^a
\rho^b$ (BL03). 

The observed cluster age distribution will obey the following
approximate power-law behaviours:

\begin{itemize}

\item ${\rm d} N_{\rm cl} / {\rm d} t \propto t^{\zeta(1-\alpha)}$
for young clusters as a result of fading;

\item ${\rm d} N_{\rm cl} / {\rm d} t \propto t^{(1-\alpha)/\gamma}$
for old clusters as a result of disruption.

\end{itemize}
Similarly, the mass spectrum of the observed clusters will be

\begin{itemize}

\item ${\rm d} N_{\rm cl} / {\rm d} M_{\rm cl} \propto M_{\rm
cl}^{(1/\zeta)-\alpha}$ for low-mass clusters as a result of fading;

\item ${\rm d} N_{\rm cl} / {\rm d} M_{\rm cl} \propto M_{\rm
cl}^{\gamma-\alpha}$ for high-mass clusters as a result of disruption.

\end{itemize}  
 
Thus, both distributions will show a double power law with slopes
determined by $\alpha$, $\zeta$ and $\gamma$. The crossing points,
$t_{\rm cross}$ and $M_{\rm cross}$, are determined by the cluster
formation rate and by the $\tdisref$ time-scale. In these expressions,
$N_{\rm cl}$ is the total number of clusters in a given sample, and so
${\rm d} N_{\rm cl} / {\rm d} M_{\rm cl}$ and ${\rm d} N_{\rm cl} / {\rm
d} t$ are the numbers of clusters per constant bin in mass and age,
respectively. 

In BL03 and de Grijs et al.  (2003b) we showed that the observed age and
mass distributions of the star cluster systems in a number of
well-studied galaxies indeed show the predicted double power-law
behaviour, where applicable modified by a non-constant (bursty) cluster
formation rate.  From the analysis of these observed distributions BL03
showed that the value of $\gamma$ is approximately constant, $(\gamma =
0.62 \pm 0.06)$, under the very different environmental conditions in
their sample galaxies, but the characteristic disruption time-scales
(i.e., $t_4^{\rm dis}$) differ significantly from galaxy to galaxy. 

In Sect. \ref{ngc3310.sect} we will derive the formation history of the
NGC 3310 cluster system; we will distinguish between star clusters in
the starburst ring and those detected outside (Sect. 
\ref{ringvsnon.sect}), and interpret our results in the context of star
cluster disruption processes and time-scales (Sect. 
\ref{ngc3310disr.sect}). We will then discuss the system of young star
clusters and star-forming regions in NGC 6745 in Sect. 
\ref{ngc6745.sect} in the context of its very recent tidal encounter
with a small companion galaxy (Sect. \ref{geometry.sect}), and explore
the cluster properties in the framework of their disruption history
(Sections \ref{n6745prop.sect} and \ref{n6745icmf.sect}). As we will
see, both cluster systems are too young to have already undergone
significant disruption, so that we are in fact observing their
properties in close to initial conditions. We will discuss the derived
slopes for the initial cluster mass functions for both galaxies in the
general context of star cluster formation in Sect. \ref{icmf.sect}, and
then summarize our results and conclusions in Sect. 
\ref{conclusions.sect}. 

\section{Measuring the initial cluster mass function slope in NGC 3310}
\label{ngc3310.sect}

In Paper I we obtained robust age and mass estimates for $\sim 150$ star
clusters in the nearby starburst galaxy NGC 3310.  We used a number of
passband combinations based on archival {\sl Hubble Space Telescope
(HST)} observations from the ultraviolet (UV) to the near-infrared (NIR)
to achieve this.  In this paper, we will use these age and mass
estimates to derive further details of the clusters' formation history
and their subsequent evolution. 

\begin{figure*}
\hspace*{0.2cm}
\psfig{figure=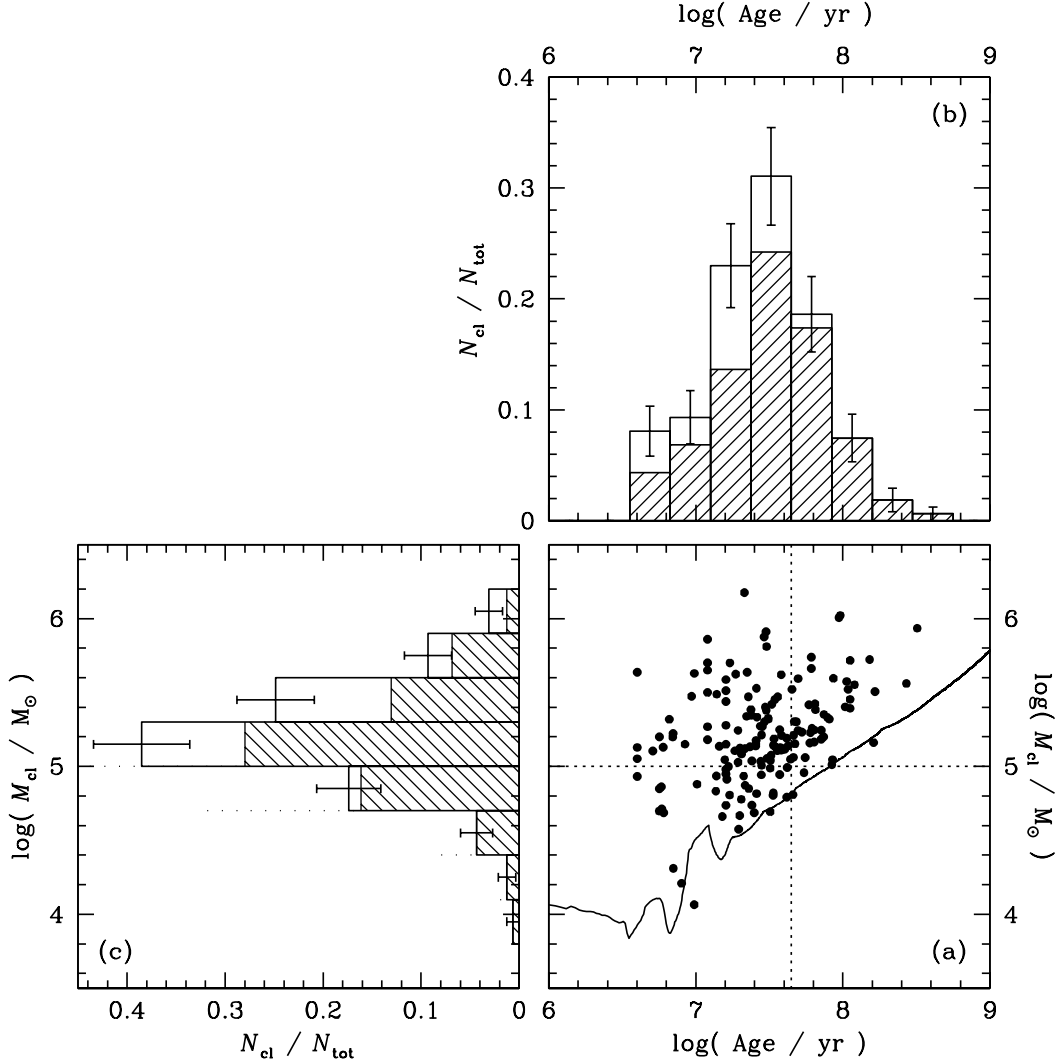,width=15cm}
\vspace*{-0.5cm}
\caption{\label{agemass1.fig}(a) -- Distribution of the NGC 3310
clusters in the (age vs.  mass) plane.  Overplotted is the expected,
age-dependent detection limit.  This model prediction is based on a
detection limit of $m_{\rm F606W} \sim V = 21$ (see text) and
$(m-M)_{{\rm NGC}3310} = 30.57$, assuming no extinction.  The features
around 10 Myr are caused by the appearance of red supergiants.  (b) and
(c) -- Distributions of, respectively, the ages and masses of the
compact clusters in the centre of NGC 3310.  The open histograms, with
their associated Poissonian error bars, represent the full cluster
sample; the shaded histogram in panel (b) corresponds to the clusters
with masses greater than $\log( M_{\rm cl}/M_\odot ) = 5.0$, indicated
by the horizontal dotted line in panel (a), and the shaded histogram in
panel (c) represents the youngest clusters, $\log( {\rm Age / yr} ) \le
7.65$, corresponding to the data points to the left of the vertical
dotted line in panel (a).  The uncertainties are of the order of the
histogram bin sizes.}
\end{figure*}

Figure \ref{agemass1.fig}a shows the distribution of the NGC 3310
clusters in the age vs.  mass plane.  The solid line overplotted on the
figure shows the expected effect of evolutionary fading of an
instantaneously formed SSP.  The fading line shown is based on zero
extinction, and applies to an observed cluster system with a limiting
magnitude of $V = 21$ at the distance of NGC 3310 ($m-M = 30.57$; see
Paper I), and our adopted, Salpeter-type IMF (with stellar masses {\it
m} in the range $0.15 \le m/M_\odot \le (50-70)$, the upper limit
depending on the metallicity and determined by the mass coverage of the
Padova isochrones), predicted by the G\"ottingen SSP models (Schulz et
al.  2002, updated to include nebular emission by Anders \& Fritze--v. 
Alvensleben 2003; see also Paper I).  The predicted lower limit agrees
very well with our data points. 

Because of the way in which we performed our cluster selection in Paper
I, our adopted photometric completeness limit is ultimately determined
by the interplay of two effects.  These include the photometric depth of
the shallowest exposure for which we required genuine source detections
(i.e., the F300W image, for which we determined an average 90\%
completeness fraction at $V \simeq 22$; see Figure 1 in Paper I) and the
further loss of completeness due to the final, visual verification of
the cluster candidates detected automatically.  We conservatively
estimate this latter step, which dominates the {\it V}-band photometry,
to reduce our sample completeness by of order one more magnitude (see
the detailed discussion on completeness effects for the analysis of the
M51 cluster system in N.  Bastian et al., in prep.), so that the
limiting magnitude at $\sim 90$\% completeness for the full sample of
NGC 3310 clusters is $V \simeq 21$. 

For a nominal extinction of $A_V = 0.2$ mag (see Paper I), the detection
limit will shift to higher masses by $\Delta \log( M_{\rm cl}/ M_\odot )
\simeq 0.08$, which is well within the uncertainties associated with our
mass determinations. 

Figures \ref{agemass1.fig}b and c show the distributions of,
respectively, the ages and masses of the compact clusters in the centre
of NGC 3310.  The uncertainties in our fit results are of the order of
the histogram bin sizes (Paper I).  The open histograms, with their
associated Poissonian error bars, represent the full cluster sample,
whereas the shaded histograms show the effects of adopting a fixed age
or mass cut-off.  The shaded histogram in Fig.  \ref{agemass1.fig}b
corresponds to the clusters with masses greater than $\log( M_{\rm
cl}/M_\odot ) = 5.0$, as indicated by the horizontal dotted line in Fig. 
\ref{agemass1.fig}a (arbitrarily chosen), and the shaded histogram in
Fig.  \ref{agemass1.fig}c represents (for an arbitrary age cut-off) the
youngest clusters, $\log( {\rm Age / yr} ) \le 7.65$, corresponding to
the data points to the left of the vertical dotted line in Fig. 
\ref{agemass1.fig}c.  The age and mass-dependent completeness limit
introduces some skewness into the age and mass distributions compared to
fixed age and mass cut-offs. 

\subsection{Starburst ring versus non-ring clusters}
\label{ringvsnon.sect}

\begin{figure*}
\hspace*{1.2cm}
\psfig{figure=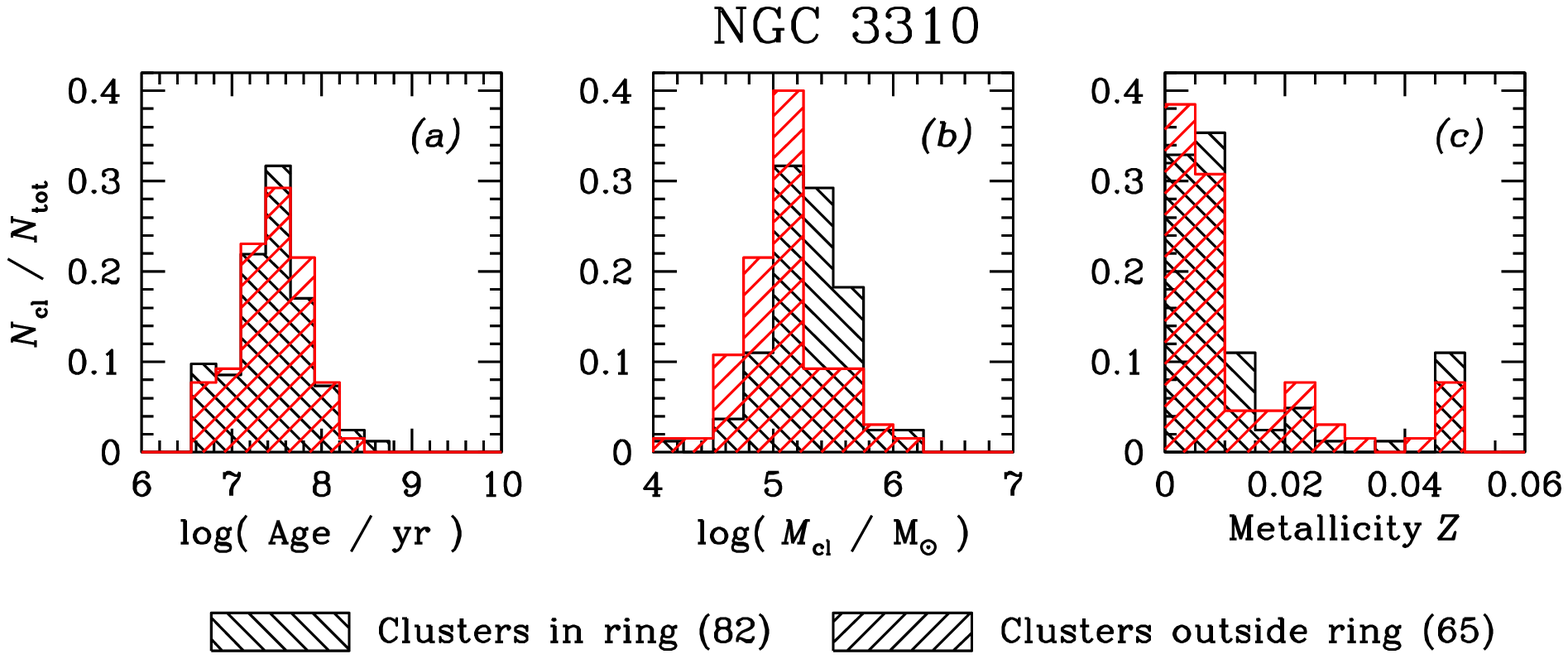,width=15cm}
\vspace*{-8.5cm}
\caption{\label{ringvsnon.fig}Comparison of the basic properties of the
ring clusters with those of the non-ring cluster sample.}
\end{figure*}

The morphology over a wide range of wavelengths (from X-rays to radio
waves) of the central regions of NGC 3310 is dominated by a bright,
relatively dense ring-like structure containing a large number of
actively star-forming regions and young star cluster candidates (see
Paper I for a review). Conselice et al. (2000) suggested a bar-driven
origin for the active starburst in the ring, combined with the recent
infall of a companion galaxy. Such a scenario provides a natural
explanation for the low metallicity observed in the star-forming knots
near the galactic centre (see Paper I for an overview), while it also
explains why we observe concentrated star formation in star clusters or
luminous H{\sc ii} regions in such a tightly-wound ring-like
structure surrounding the centre (e.g., van der Kruit \& de Bruyn 1976,
Telesco \& Gatley 1984, Pastoriza et al. 1993, Meurer et al. 1995,
Smith et al. 1996, Conselice et al. 2000, Elmegreen et al. 2002),
coinciding with the end of the nuclear bar (Conselice et al. 2000, but
see D\'\i az et al. 2000), but not inside this ring. 

\subsubsection{A comparison of basic properties}
\label{basicprop.sec}

We identified 82 of the 147 star clusters in our sample to coincide with
this circumnuclear ring, with the 65 remaining objects being located
{\it outside} this starburst ring.  In Fig.  \ref{ringvsnon.fig} we
compare the basic properties of these cluster samples.  The peaks in the
age distributions are observed at $\log( {\rm Age / yr} ) = 7.45$ and
7.43 for the clusters in and outside the ring, respectively, with
corresponding Gaussian sigmas, $\sigma_{\rm G}$, of 0.40 and 0.38 in
logarithmic age space.  However, while the age (and also the
metallicity) distributions of both cluster samples are statistically
indistinguishable, the ring clusters appear to peak at slightly greater
masses than those found outside the ring: the mass distribution of the
clusters in the ring peaks at $\log( M_{\rm cl}/M_\odot )_{\rm ring} =
5.29$, versus $\log( M_{\rm cl}/M_\odot )_{\rm non-ring} = 5.13$ for the
clusters outside the ring.  Their Gaussian sigmas are also significantly
different, at $\sigma_{\rm G,ring} = 0.32$ and $\sigma_{\rm G,non-ring}
= 0.51$, respectively. 

We examined whether this difference could be due to a spatial dependence
of the completeness fraction, in the sense that one might expect a lower
completeness in the ring area with its higher background flux than
outside the ring.  This difference in the mean (logarithmic) mass
corresponds to a difference in the peak of the cluster luminosity
function (CLF), in any passband, of $\Delta m_{\rm peak} \simeq 0.4$
mag, and the difference in the Gaussian sigmas of the distributions is
also in the sense expected if this were due to different and variable
levels of completeness.  We examined the mean surface brightness levels
in and outside the starburst ring, respectively, both in the F606W image
and in the slightly shallower F300W passband.  The difference in the
mean surface brightness levels between the starburst ring and the area
outside the ring is $\Delta (\Sigma_{\rm non-ring} - \Sigma_{\rm ring})
\sim 0.3-0.9$ mag arcsec$^{-2}$ and $\sim 0.1-0.8$ mag arcsec$^{-2}$ in
the F300W and F606W images, respectively, based on a per-pixel
comparison. 

The overabundance of low-mass clusters outside the ring compared to
their counterparts in the starburst ring might therefore (at least
partially) be due to a spatial dependence of the completeness fraction. 
However, there is a clear and significant excess of higher-mass clusters
($5.25 \lesssim \log( M_{\rm cl}/M_\odot ) \lesssim 5.75$) {\it in} the
ring compared to the non-ring cluster sample, even after taking the
systematic uncertainties in our mass estimates into account (the
estimated $\sim 1 \sigma$ systematic uncertainties are of the order of
the histogram bin sizes).  If both the ring and the non-ring cluster
populations were drawn from the same parent population, one would not
expect to observe such large differences, and thus it appears that the
physical conditions in the starburst ring, such as caused by the effects
of the proposed bar-driven instabilities, the higher density of the
interstellar medium (ISM), and the associated higher likelihood to
encounter significant propagating shock waves, may be conducive for the
formation of higher-mass star clusters, on average, than in the
relatively more quiescent environment of the main galactic disc.  Thus,
the origin of the differences in the global mass distributions at the
high-mass end of the two cluster samples is most likely found in their
respective ISM properties. 

If the cluster formation in the starburst ring were predominantly due to
the bar-driven instabilities suggested by Conselice et al.  (2000), one
would, to first order, expect a narrower age distribution for the
clusters in the ring with respect to that of the non-ring clusters in
the general field of the central galactic disc.  However, both age
distributions are characterized by a median age of $\langle \log( {\rm
Age/yr} ) \rangle \simeq 7.5$ ($\sim 30$ Myr), and an age spread of
$\sigma_{\rm G}(\log[ {\rm Age/yr} ] ) \simeq 0.4$ ($\sim 70$ Myr, from
$\sim 80 - 10$ Myr ago).  Furthermore, if we examine the age
distributions of the clusters in linear age space, it appears that
clusters both in and outside the starburst ring have been forming
approximately constantly, but at a $\sim 2-3 \times$ higher level during
the past $\sim 40$ Myr than before.  Thus, we conclude that the
starburst producing the NGC 3310 ring clusters has been ongoing for at
least the past 40 Myr, at an approximately constant cluster formation
rate during the burst period.  Star cluster formation has proceeded at a
similar rate in the general central disc of the galaxy; due to the
extreme youth of both samples of starburst-induced star clusters,
statistically significant differences between the age distributions of
the ring and non-ring populations have not yet had time to develop,
which implies that we are still observing at least part of the cluster
sample in the environment determined by their initial conditions. 

\subsubsection{Effects of metallicity changes on the systematic
uncertainties} 

Finally, we caution that the peak of cluster formation, while relatively
robust (see, e.g., Paper I for a discussion), is undoubtedly broadened
by the various uncertainties entering the age-dating process.  Some of
these (systematic) uncertainties were discussed in detail in Paper I,
where we concentrated our discussion on the choice of passband
combination for this type of analysis.  Here, we will try to obtain a
further handle on the systematic uncertainties.  If the cluster
formation in the starburst ring were induced by bar-driven instabilities
acting on very short time-scales (i.e., on time-scales short compared to
the median age of the cluster sample), one would not expect the
resulting cluster sample to be characterized by a large range in
metallicity.  This argument holds both if the progenitor gas clouds
originated from the main NGC 3310 gas reservoir, and also if they came
from a low-metallicity dwarf galaxy that was cannibalized (assuming that
such a galaxy would exhibit a fairly small range in metallicities, as
suggested for this particular interaction; see Paper I).  The former
case is further strengthened by observations in other galaxies that old
galactic discs generally show smooth, very slowly declining radial
metallicity gradients, such that at any given small distance range from
the galactic centre (e.g., as in the NGC 3310 starburst ring) the
stellar metallicity is approximately constant (for the Galaxy see, e.g.,
Twarog et al.  1997, and the discussion therein).  This is to some
extent already reflected in Fig.  \ref{ringvsnon.fig}c, where we derived
that $> 80$\% of the clusters in the ring have metallicities $Z <
0.015$. 

\begin{figure}
\psfig{figure=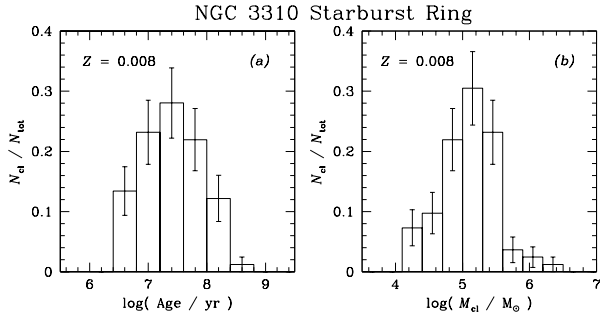,width=8.5cm}
\vspace*{-4.3cm}
\caption{\label{ringz3.fig}Age and mass distributions of the clusters in
the NGC 3310 starburst ring, obtained under the assumption that {\it
all} clusters have a fixed metallicity of $Z=0.008$.}
\end{figure}

We will now assume a mean metallicity of $Z = 0.008$ for the clusters in
the starburst ring, and rederive their age and mass distributions.  This
will give us a further handle on the systematic, model-dependent
uncertainties entering the age-dating and mass-estimation process.  In
Fig.  \ref{ringz3.fig} we show the age and mass distributions obtained
under the assumption that the ring clusters all have a fixed metallicity
of $Z = 0.008$.  The peaks in the age and mass distributions occur at
$\langle \log( {\rm Age/yr} ) \rangle = 7.34$ and $\langle \log( M_{\rm
cl}/M_\odot ) \rangle = 5.09$, respectively, while the distributions are
characterized by Gaussian sigmas $\sigma_{\rm G,age} = 0.50$ and
$\sigma_{\rm G,mass} = 0.42$.  As a reminder, the age and mass
distributions resulting from retaining the metallicity as a free
parameter (Sect.  \ref{basicprop.sec}), were $\langle \log( {\rm Age/yr}
) \rangle \simeq 7.45$ and $\langle \log( M_{\rm cl}/M_\odot ) \rangle =
5.29$, respectively, with $\sigma_{\rm G,age} = 0.40$ and $\sigma_{\rm
G,mass} = 0.32$.  We see that while the peak values are retained with a
reasonable robustness (although there is, as expected, some effect
caused by the age--metallicity degeneracy), both distributions have
broadened significantly.  This confirms, therefore, that (i) it is very
important to determine all of the free parameters (age, metallicity, and
extinction, and the corresponding mass estimates) {\it independently}
for each individual cluster, instead of assuming a generic value for any
of these parameters (if a sufficient number of broad-band passbands
covering a large wavelength range are available); and (ii) the widths of
the distributions (assuming a fixed metallicity) are broader than their
intrinsic widths owing to the propagation of model and measurement
uncertainties (e.g., de Grijs et al.  2003b). 

For comparison, we also computed the characteristics of the age and mass
distributions for the NGC 3310 clusters outside the ring, under the same
assumption of a fixed metallicity of $Z=0.008$ for all clusters.  The
trend between these new determinations and those of the
fixed-metallicity results for the ring clusters remains similar as for
the situation in which we allowed the metallicity to be an additional
free parameter: the age distribution peaks again at a similar median
age, $\langle \log( {\rm Age/yr} ) \rangle_{\rm non-ring} = 7.37$, with
a Gaussian sigma of $\sigma_{\rm G,non-ring} = 0.52$.  The median of the
mass distribution occurs at a significantly lower mass than for the ring
clusters, $\langle \log( M_{\rm cl}/M_\odot ) \rangle_{\rm non-ring} =
4.93$, with a Gaussian sigma of $\sigma_{\rm G,non-ring} = 0.60$.  Thus,
we arrive at similar conclusions as before for the mass distribution
of the non-ring compared to
the ring sample of clusters in NGC 3310. 

\subsection{The star cluster disruption context}
\label{ngc3310disr.sect}

\begin{figure*}
\hspace*{1.2cm}
\psfig{figure=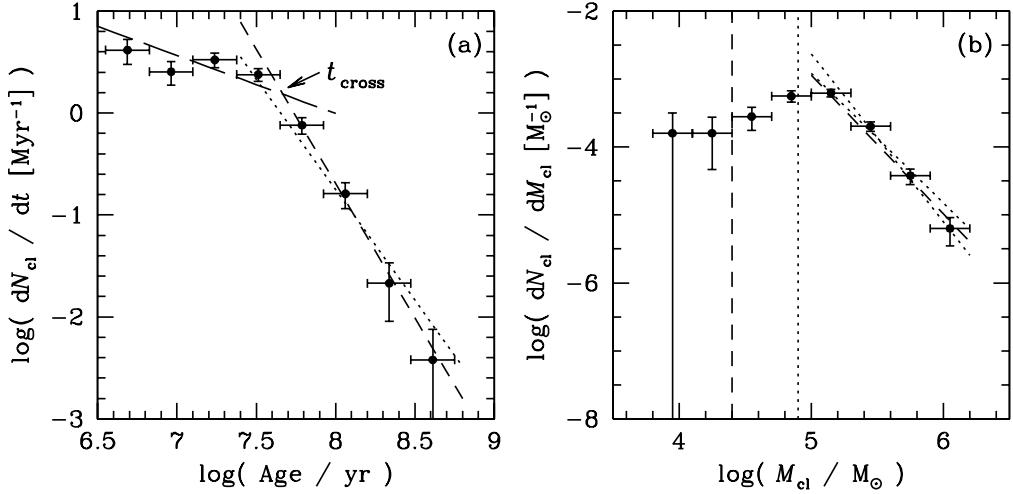,width=15cm}
\vspace*{-7.5cm}
\caption{\label{timescales1.fig}(a) -- The cluster formation rate (in
number of clusters per Myr) in NGC 3310 as a function of age. The
long-dashed line for $\log( {\rm Age / yr} ) \le 8$ is the expected
effect of a fading SSP, for a constant ongoing cluster formation rate,
shifted vertically to best match the data points for $\log( {\rm Age /
yr} ) \le 7.5$. The short-dashed and dotted lines are the best-fitting
and maximum deviating power-law slopes, as described in the text. (b)
-- Mass spectrum of the NGC 3310 clusters (number of clusters per mass
bin); line encoding as in panel (a). The vertical dashed line at $\log(
M_{\rm cl} / M_\odot ) = 4.4$ corresponds to our completeness limit, for
a cluster age of $\log( {\rm Age / yr} ) = 7.5$, while the vertical
dotted line corresponds to the same completeness limit at an age of
$\log( {\rm Age / yr} ) = 8.0$.}
\end{figure*}

In Fig. \ref{timescales1.fig} we show the formation rate and the mass
spectrum of the entire NGC 3310 cluster sample. These distributions
depend on the cluster formation history and on the cluster disruption
time-scale governing the centre of NGC 3310. 

For our analysis of the cluster disruption history in the centre of NGC
3310 we adopt a slope $\zeta=0.648$ for the evolutionary fading of
clusters in the {\it V} band, derived from theoretical SSP models (see
also BL03, de Grijs et al.  2003b).  For the mass scaling of the
disruption time-scale we adopt $\gamma=0.62$ (BL03, de Grijs et al. 
2003b). 

We also assume that the cluster formation rate has been approximately
constant during the periods of interest for our analysis, $\log( {\rm
Age/yr} ) \lesssim 7.5$ and $\log( {\rm Age/yr} ) \gtrsim 7.5$. As
shown in Sect. \ref{ringvsnon.sect}, this assumption seems justified,
within the uncertainties. 

\begin{figure*}
\hspace*{1.2cm}
\psfig{figure=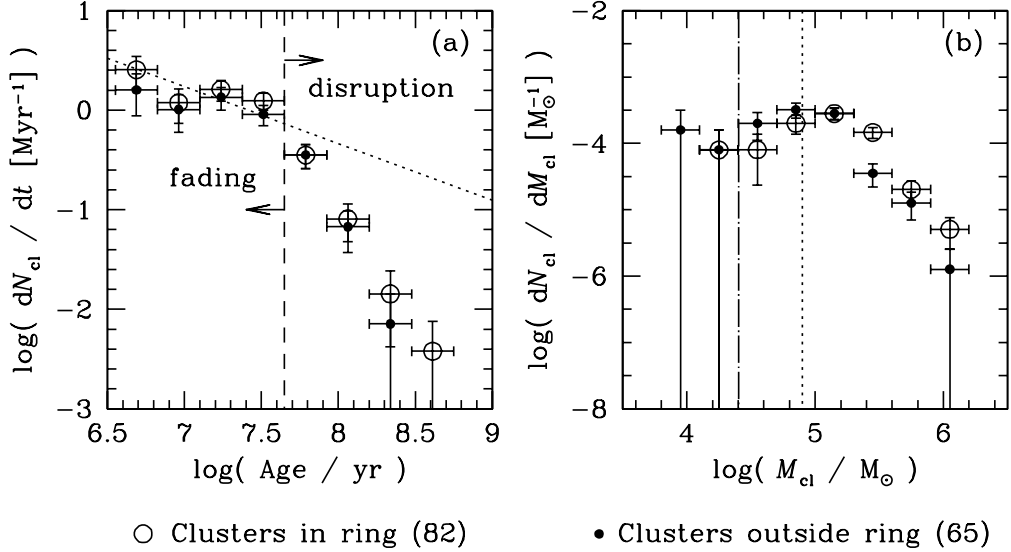,width=15cm}
\vspace*{-6.5cm}
\caption{\label{timescales2.fig}Comparison of the cluster formation
rates and mass spectra of the clusters in vs. outside the NGC 3310
starburst ring. The age ranges in which stellar evolutionary fading and
cluster disruption dominate are indicated; the vertical dash-dotted line
in panel (b) indicates our completeness limit for a cluster age of
$\log( {\rm Age/yr} ) = 7.5$, while the vertical dotted line corresponds
to the same completeness limit, but at an age of $\log( {\rm Age/yr} ) =
8.0$.} 
\end{figure*}

In Fig.  \ref{timescales1.fig}a, the long-dashed line for the youngest
ages, $\log( {\rm Age / yr} ) \le 8$, is the expected effect of a fading
SSP (in the {\it V} band), for a constant ongoing cluster formation
rate, shifted vertically to best match the data points for $\log( {\rm
Age / yr} ) \le 7.5$.  The short-dashed lines in both Figs. 
\ref{timescales1.fig}a and b are the best-fitting power-law slopes,
obtained for the age and mass ranges $\log( {\rm Age/yr} ) \ge 7.7$ and
$\log( M_{\rm cl}/M_\odot ) \ge 5.1$, while the dotted lines are
indicative of the likely uncertainties in the slopes.  The latter were
obtained from subsets of the data points used for the dashed fits,
determined over the full ranges.  Our results are robust with respect to
changes in the adopted bin size in $\log( {\rm Age/yr} )$ and $\log(
M_{\rm cl}/M_\odot )$.  The vertical dashed line in Fig. 
\ref{timescales1.fig}b at $\log( M_{\rm cl} / M_\odot ) = 4.4$
corresponds to our completeness limit, for a cluster age of $\log( {\rm
Age / yr} ) = 7.5$.  However, since our cluster sample is characterized
by a range of ages up to $\sim 10^8$ yr, it follows that incompleteness
plays at least some role for masses up to $\log( M_{\rm cl}/M_\odot )
\sim 4.9$ (see the vertical dotted line in Fig. 
\ref{timescales1.fig}b).  For the age distribution the effects of
incompleteness are less complicated, since they are determined solely by
(the extension of) the fading line in Fig.  \ref{timescales1.fig}a. 
Moreover, the effects of the non-constancy of the cluster formation rate
are less transparent and therefore harder to disentangle from the mass
distribution than from the age distribution (see de Grijs et al.  2003b
for a discussion). 

The short-dashed line in Fig.  \ref{timescales1.fig}a, which shows the
effect of cluster disruption, has a predicted slope of
$(1-\alpha)/\gamma$ and an observed slope of $-2.64 (\pm 0.14) \pm
0.47$, where the first uncertainty represents the formal uncertainty in
the fit and the second the likely uncertainty resulting from our choice
of fitting range (i.e., the dotted fits in Figs.  \ref{timescales1.fig}a
and b).  If we adopt a cluster IMF slope of $\alpha=2.0$ (see Sect. 
\ref{icmf.sect}) we find that the slope of the cluster disruption law is
$\gamma \simeq 0.38 - 0.46$, with $ t_{\rm dis} \propto M_{\rm
cl}^\gamma$.  This value is smaller than the mean value of $\gamma
\simeq 0.62$ found by BL03 from cluster samples in different galaxies. 
The crossing point of the two power-law fits is at $\log t_{\rm cross}
\simeq 7.7$.  Substituting this value in Eq.  (15) of BL03, we find that
the characteristic disruption time of a $10^4 M_\odot$ cluster is $\log
t_4^{\rm dis} = 7.9$.  This implies that the disruption time-scale of
clusters with an initial mass greater than $10^5 M_\odot$ will be
greater than $2 \times 10^8$ yr.  The slope of the mass distribution in
Fig.  \ref{timescales1.fig}b, for clusters more massive than $10^5
M_\odot$ is $-2.04 (\pm 0.23)^{+0.13}_{-0.43}$, where the first
uncertainty represents the formal uncertainty in the fit and the second
the likely uncertainty resulting from our choice of fitting range.  All
these clusters are younger than $2 \times 10^8$ yr.  (In fact, all our
sample clusters, except one, are younger than this age.) Thus,
disruption has not yet affected the mass distribution of these massive
clusters and hence the observed slope of the mass distribution is the
slope of the cluster IMF.  We note that this new value of the NGC 3310
cluster IMF slope closely matches both our earlier CLF slope
determination of $\alpha = 1.8 \pm 0.4$ obtained from the cluster mass
function in Paper I, and the best estimate of Elmegreen et al.  (2002),
$2.2 \lesssim \alpha \lesssim 2.4$. 

Finally, in Fig.  \ref{timescales2.fig} we compare the clusters in the
ring with those outside the ring.  Within the observational
uncertainties, we do not detect any differences in the slopes of either
the age or the mass distributions.  This is another argument
highlighting the very young age of this cluster system compared to the
expected characteristic disruption time-scale, even in the higher
density environment of the starburst ring. 

\section{Star formation in the Bird's Head Galaxy}
\label{ngc6745.sect}

\subsection{Observations and data preparation}

Optical observations of NGC 6745 in the minimum of four passbands
spanning the entire wavelength range from the {\it U} (F336W) to the
{\it I} band (F814W) were obtained between 18 and 20 March 1996, as part
of {\sl HST} GO programme 6276 (PI Westphal), using the Wide Field
Planetary Camera 2 ({\sl WFPC2}). Multiple exposures were obtained
through each filter to facilitate the removal of cosmic rays; the
galactic centre was placed on the WF3 chip (pixel size 0.0997 arcsec) in
all cases. An overview of the available {\sl HST} observations is given
in Table \ref{obs.tab}. 

\begin{table}
\caption[ ]{\label{obs.tab}Overview of the {\sl HST} observations of NGC
6745
}
{\scriptsize
\begin{center}
\begin{tabular}{crc}
\hline
\hline
\multicolumn{1}{c}{Filter} & \multicolumn{1}{c}{Exposure time} &
\multicolumn{1}{c}{ORIENT$^a$} \\
& \multicolumn{1}{c}{(sec)} & \multicolumn{1}{c}{($^\circ$)} \\  
\hline
F336W & 2400, 7$\times$2800 & 145.52 \\
F555W & 2$\times$1100       & 145.52 \\
      & 2$\times$1300       & 145.52 \\
F675W & 4$\times$1300       & 145.52 \\
F814W & 4$\times$1300       & 145.52 \\
\hline
\end{tabular}
\end{center}  
{\sc Note:} $^a$ -- 
Orientation of the images (taken from the image header), measured North
through East with respect to the V3 axis (i.e., the X=Y diagonal of the
WF3 CCD $+ 180^\circ$). 
}
\end{table} 

Pipeline image reduction and calibration of the {\sl WFPC2} images were
done with standard procedures provided as part of the {\sc
iraf/stsdas}\footnote{The Image Reduction and Analysis Facility (IRAF)
is distributed by the National Optical Astronomy Observatories, which is
operated by the Association of Universities for Research in Astronomy,
Inc., under cooperative agreement with the National Science Foundation.
{\sc stsdas}, the Space Telescope Science Data Analysis System, contains
tasks complementary to the existing {\sc iraf} tasks. We used Version
2.2 (August 2000) for the data reduction performed in this paper.}
package, using the updated and corrected on-orbit flat fields and   
related reference files most appropriate for the observations.

We obtained source lists and, subsequently, source photometry for all of
our NGC 6745 observations following procedures identical to the ones
described in detail in Paper I. For NGC 6745, at a distance $D \simeq
68$ Mpc ($m-M = 34.17$; assuming a systemic velocity $v_{r,\odot} = 4545
\pm 60$ km s$^{-1}$ [Falco et al. 1999], a 208 km s$^{-1}$ correction
for infall of the Local Group towards the Virgo cluster, and $H_0 = 70$
km s$^{-1}$ Mpc$^{-1}$), all compact clusters and most star forming
regions appear as point-like sources. The final list of verified cluster
candidates detected at at least four times the background r.m.s. noise
level in all four passbands contains 177 objects. 

\begin{figure}
\psfig{figure=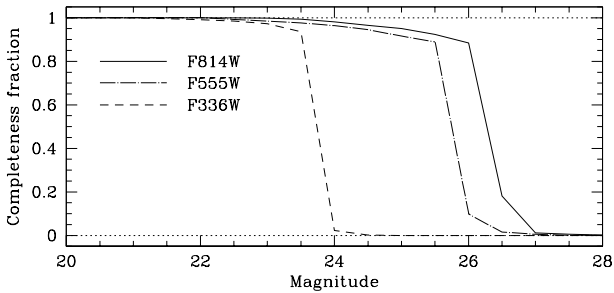,width=8.5cm}
\vspace{-4cm}
\caption{\label{compl6745.fig}Formal completeness curves for the full {\sl
WFPC2} field of view of the NGC 6745 observations. The different line
styles refer to different passbands, as indicated in the figure.}
\end{figure}

We estimated the completeness of our source lists by using synthetic
source fields consisting of PSFs.  We created artificial source fields
for input magnitudes between 20.0 and 28.0 mag, in steps of 0.5 mag,
independently for each of the F336W, F555W, and F814W passbands.  We
then applied the same source detection routines used for our science
images to the fields containing the combined galaxy image and the
artificial sources.  The results of this exercise are shown in Fig. 
\ref{compl6745.fig}.  These formal completeness curves were corrected
for the effects of blending or superposition of multiple randomly placed
artificial PSFs as well as for the superposition of artificial PSFs on
genuine objects.  A detailed description of the procedures employed to
obtain these completeness curves was given in Paper I. 

We found that the effects of image crowding are small: only $\lesssim
1.5-2.5$\% of the simulated objects were not retrieved due to crowding,
either in the artificial or in the combined frames.  However, the
effects of the bright and irregular background and dust lanes are large,
resulting in variable completeness fractions across the galaxy images. 
As a general rule, however, the curves in Fig.  \ref{compl6745.fig} show
that the formal completeness drops below $\sim 90$\% for F336W $\gtrsim
23.5$ mag, for F555W $\gtrsim 25.5$ mag, and for F814W $\gtrsim 26$ mag;
the actual completeness limits are likely somewhat shallower, resulting
from our selection criteria (cf.  N.  Bastian et al., in prep.). 

Foreground stars are not a source of confusion.  The standard Milky Way
star count models (e.g., Ratnatunga \& Bahcall 1985) predict roughly
1--2 foreground stars for the equivalent standard filter of F555W
$\lesssim 24$ in our field of view.  Background objects may pose a
(small) problem, however, in particular among the fainter sources, since
we did not impose any roundness or sharpness constraints on our extended
source detections, in order not to omit unrelaxed young clusters and
star-forming regions from our final sample.  However, such objects
should be easily identifiable once we have obtained aperture photometry
for our complete source lists, as they are expected to have
significantly different colours.  Background galaxies at redshifts
greater than about 0.1 are expected to have extremely red $(m_{\rm
F336W}-m_{\rm F814W})$ colours compared to their local counterparts and
the young star clusters and star forming regions expected in NGC 6745. 

We will discuss the derived parameters and their implications for the
evolution of the galaxy's star cluster system in Sects. 
\ref{n6745prop.sect} and \ref{n6745icmf.sect}, respectively. 

Additionally, H{\sc i} 21cm observations (at $\sim 1.399$ GHz) were
obtained with all 27 antennae of the Very Large Array (VLA) in C
configuration, on 28 May 2000, with an on-source exposure time of 24765
s ($\sim 7$ hr).  Primary calibrators were observed at the beginning and
end; secondary (phase) calibrator integrations were uniformly
interspersed throughout the run.  The {\it (u,v)} plane coverage was
almost perfectly isotropic, and 8192 CLEAN cycles (with uniform {\it
(u,v)} weighting) yielded an essentially symmetric $\sim 13.75$ arcsec
FWHM beam.  The 31 $\sim 42.5$ km s$^{-1}$ channels were centered on
4541 km s$^{-1}$ (channel 16), with channels 11 through 21 containing
all of the H{\sc i} emission.  Continuum subtraction was accomplished in
{\it (u,v)} space.  We will discuss these observations in detail in
Sect.  \ref{geometry.sect}. 

\subsection{The interaction geometry and its consequences}
\label{geometry.sect}

\begin{figure*}
\hspace*{0.2cm}
\psfig{figure=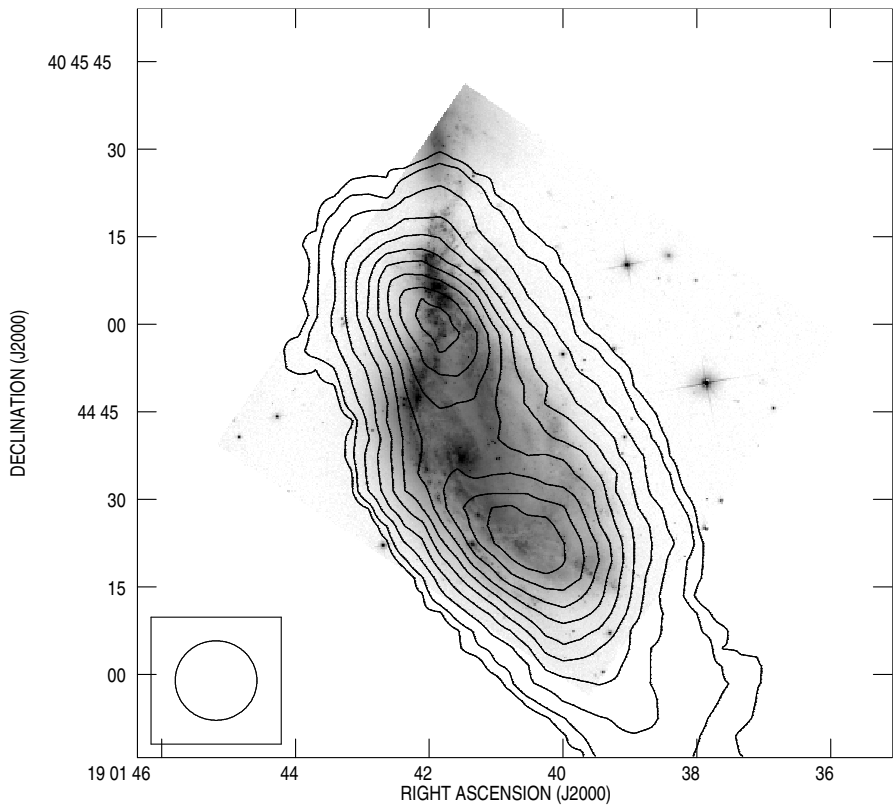,width=12cm,angle=-90}
\caption{\label{n6745HI0.fig}NGC 6745 H{\sc i} distribution shown as
contours superimposed on a negative logarithmic reproduction of the
F555W {\sl HST}/WFPC2 image of the galaxy.  The VLA/C beam size is 13.75
arcsec (FWHM) and is indicated in the lower left corner.  Contour
intervals are 10 per cent of the maximum flux density of the southern
maximum, $9.565 \times 10^2$ Jy beam$^{-1}$ m s$^{-1}$, except for the
lowest contour, which is at 5 per cent of the maximum.  Note the
East-West asymmetry, the stretching of the lowest contours to the South,
and the absence of detectable emission from the northern galaxy. For
reference, the (J2000) centre coordinates of NGC 6745a and c are R.A. =
$19^{\rm h}01^{\rm m}41.8^{\rm s}$, Dec = $40^\circ44'40''$ and R.A. =
$19^{\rm h}01^{\rm m}42.0^{\rm s}$, Dec = $40^\circ45'35''$,
respectively.} 
\end{figure*}

\begin{figure*}
\hspace*{0.2cm}
\psfig{figure=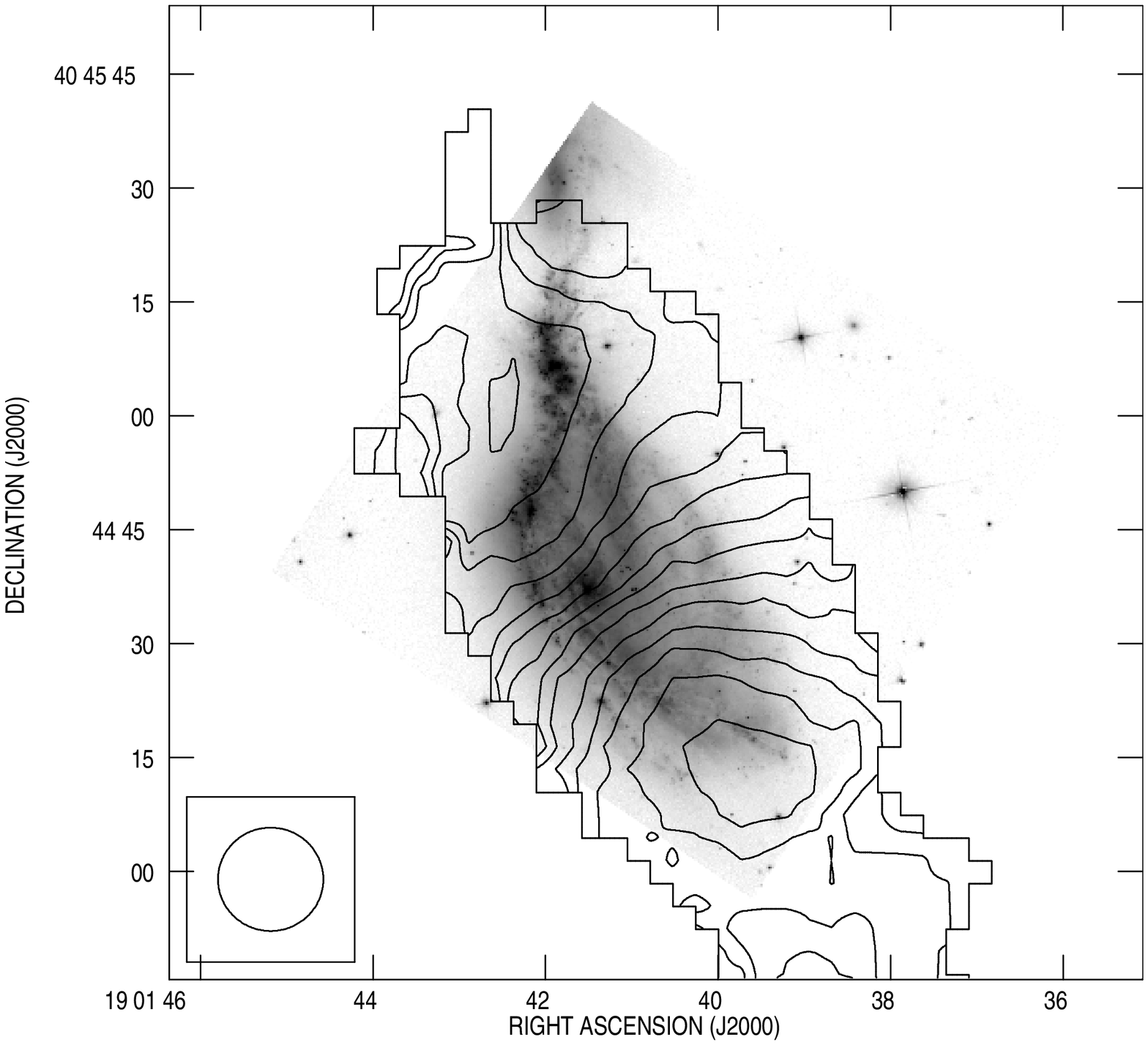,width=12cm,angle=-90}
\caption{\label{n6745HI1.fig}NGC 6745 H{\sc i} radial velocity
distribution shown as contours superimposed on a negative logarithmic
reproduction of the F555W {\sl HST}/WFPC2 image of the galaxy.  The
VLA/C beam size is 13.75 arcsec (FWHM) and is indicated in the lower
left corner.  Contours are at intervals of 20 km s$^{-1}$ beginning at
4400 km s$^{-1}$ for the smallest entire contour South West of the
center enclosing the minimum of 4388 km s$^{-1}$ and ending with the
smallest entire contour North East of the center enclosing the maximum of
4661 km s$^{-1}$.  The ``boxy'' outer line encloses map data $>8.14
\times 10^{-4}$ Jy beam$^{-1}$ (3 $\sigma$) used in the moment
calculations.  The contours exhibit, in general, the splayed structure
that is characteristic of rotating disks common in late-type spiral
galaxies.}
\end{figure*}

The optical appearance of the NGC 6745 system, and in particular the
locations of the numerous bright blue star forming complexes, are
suggestive of a tidal passage by the small northern companion galaxy
(NGC 6745c; nomenclature from Karachentsev, Karachentseva \&
Shcherbanovskii 1978) across the eastern edge of the main galaxy, NGC
6745a. The high relative velocities of the two colliding galaxies
likely caused ram pressures at the surface of contact between the
interacting interstellar clouds in both galaxies, which -- in turn --
are responsible for the triggering of enhanced star formation, most
notably in the interaction zone in between the two galaxies, NGC 6745b. 

We analysed our VLA H{\sc i} observations with as principal aim to
assess whether we could confirm this scenario, at least qualitatively,
and if we could, therefore, derive the interaction time-scale from our
cluster analysis.  The interaction-induced cluster formation scenario is
at least qualitatively supported by the H{\sc i} zeroth-moment map of
the system, shown overlaid on the optical {\sl WFPC2} image in Fig. 
\ref{n6745HI0.fig}, and the H{\sc i} velocity (first moment) map of Fig. 
\ref{n6745HI1.fig}.  First, we point out that there is no evidence for
H{\sc i} emission from the northern galaxy.  This is consistent with the
absence of significant recent star formation observed in this component
(but see Sect.  \ref{n6745fixed.sect}). 

The slight distortions in the symmetry of the overall H{\sc i}
distribution along and perpendicular to the direction of elongation may,
in fact, be consistent with tidal effects by a small, low-mass intruder
(i.e., NGC 6745c).  At the northern end of the H{\sc i} distribution the
relevant time-scales are likely so short that we would not expect to see
any significant amounts of tidally stretched H{\sc i} gas, which seems
to be the case.  At the southern end of the system, on the other hand,
where the time elapsed since the alleged South-to-North passage of the
intruder is much longer, we see the expected (asymmetric) extension of
the lower contours of the H{\sc i} distribution, as well as a high
velocity of approach at the very bottom of our field of view, which may
be qualitatively consistent with a trajectory of approach for the
intruder.  This interpretation is quantitatively supported by CO$(1-0)$
line observations of this system by Zhu et al.  (1999), who find a
similar velocity gradient between the southern component and the centre
of the main galaxy. 

The velocity data (Fig.  \ref{n6745HI1.fig}) were fit with a Brandt
function (Brandt 1960, Brandt \& Belton 1962), weighting the velocities
with the square of the intensity data (Fig.  \ref{n6745HI0.fig}).  The
residual map showed excellent agreement over all but the
lowest-intensity regions.  The best-fitting parameters include (i)
systemic velocity, $v_{r,\odot} = 4528$ km s$^{-1}$ at 4 arcsec West and
2 arcsec South of the optical nucleus; (ii) position angle of the line
of nodes = 36 degrees; (iii) inclination of the fundamental plane of the
galaxy $i = 47$ degrees; and (iv) half-amplitude value = 185 km s$^{-1}$
reached at $\sim 32$ arcsec from the location of the kinematic centre. 
The sense of the rotation of the H{\sc i} means that the perturbing
galaxy, NGC 6745c, is in a prograde orbit, which favours a strong and
rapid gravity-wave response in the affected galaxy, NGC 6745a. 

VLA continuum observations at 1.425 GHz (Condon et al.  1996) show a
more compact morphology, which more closely follows the system's stellar
distribution and far-infrared appearance (IRAS; e.g., Bushouse et al. 
1988, Condon et al.  1995, Sanders et al.  1995).  As opposed to our
$\sim 1.4$ GHz line observations that predominantly trace the cold H{\sc
i} component, the continuum flux has a non-thermal origin, as evidenced
by the $1.4 - 5$ GHz spectral index of between +0.75 and +0.83 (Condon
et al.  1991, 1995). 

\subsection{Properties of the star-forming regions}
\label{n6745prop.sect}

\begin{figure*}
\hspace*{1.2cm}
\psfig{figure=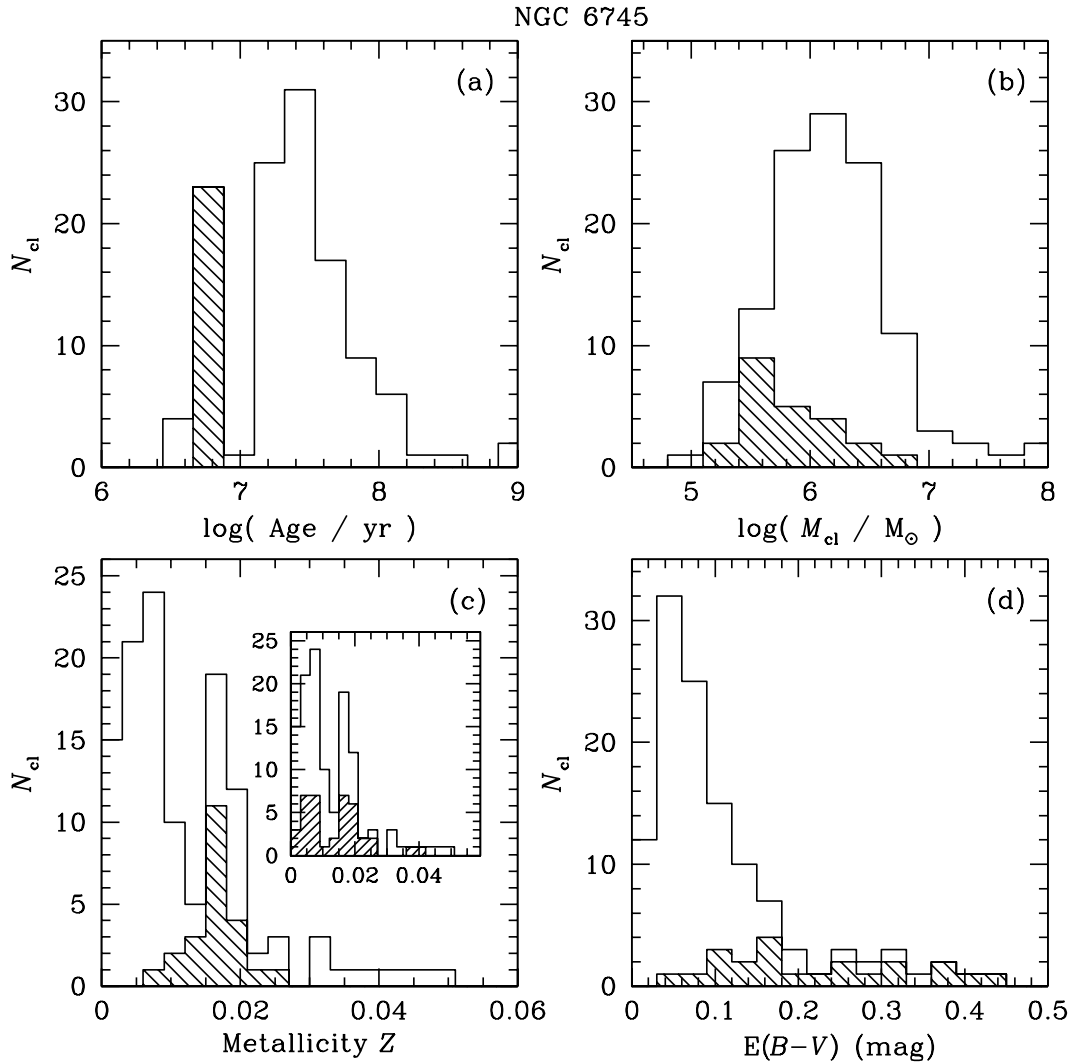,width=15cm}
\vspace*{-0.5cm}
\caption{\label{n6745data.fig}Age, mass, metallicity and extinction
distributions of the clusters in NGC 6745, based on our 4-passband
fitting technique used to determine age, metallicity and extinction
estimates simultaneously. Although the main properties of these
distributions are reasonably correct, the shaded histograms are affected
significantly by the age--metallicity degeneracy: the shaded histograms,
corresponding to the young-age peak in panel (a), show a distinct second
peak at close-to-solar metallicity (panel c), compared to the other
clusters that peak at much lower metallicities. The shaded histogram in
the inset in panel (c) is the metallicity distribution of the clusters
in the most actively star forming region connecting the main galaxy with
its smaller companion.}
\end{figure*}

Since the morphology and appearance of the NGC 6745 system are
consistent with the scenario favouring a very recent tidal encounter and
its associated triggered star and star cluster formation, we will now
use the main properties of the star clusters and star forming complexes,
such as their ages, masses and metallicities, to derive boundary
conditions for the physical conditions governing such minor merger
events.

With the lessons learnt from Paper I in mind, we applied a similar
three-dimensional $\chi^2$ minimisation (with respect to the Anders \&
Fritze--v.  Alvensleben [2003] models) to the spectral energy
distributions (SEDs) of our star cluster candidates and star forming
complexes to obtain the most likely combination of age, mass,
metallicity {\it Z} and internal extinction E$(B-V)$ (assuming a
Calzetti et al.  [2000] starburst galaxy-type extinction law for the
internal extinction) for each individual object.  Galactic foreground
extinction towards NGC 6745 was taken from Schlegel et al.  (1998).  The
resulting age, mass, metallicity and extinction distributions for the
NGC 6745 objects are shown in Fig.  \ref{n6745data.fig}. 

\subsubsection{Effects of the age--metallicity degeneracy}

Based on the knowledge gained in Paper I, we suspected that the strong
peak at young ages ($\log( {\rm Age/yr} ) \sim 6.8$; see Fig. 
\ref{n6745data.fig}a) might be an artefact caused by the
age--metallicity degeneracy. Therefore, we display these particular
clusters as the shaded histograms in Figs. \ref{n6745data.fig}b, c and
d. It is clear from an examination of the NGC 6745 cluster metallicity
distribution in Fig. \ref{n6745data.fig}c, that the clusters in the
young-age peak indeed compose a higher-than-average metallicity peak,
thus highlighting the caution required to avoid confusion arising from
this degeneracy. As a consequence of the underestimated ages for the
clusters affected by the age--metallicity degeneracy, their masses are
also biased towards lower-than-average masses. If we now turn this
argument around, and examine the properties of the clusters in the
second metallicity peak at $0.015 \lesssim Z \lesssim 0.020$, we find
that those clusters do indeed dominate the young-age peak (as expected;
not shown), and are generally biased towards younger ages overall. This
confirms the importance of the age--metallicity degeneracy for the
broad-band filters used in this paper, and for the young ages of the
star clusters and star-forming complexes in NGC 6745. The mass and
extinction properties of the clusters in this secondary metallicity peak
are consistent with those of the full cluster sample. 

\subsubsection{Case for a fixed mean metallicity}
\label{n6745fixed.sect}

In the inset of Fig. \ref{n6745data.fig}c we show the metallicity
distribution of the entire cluster sample once again, with the clusters
from the interaction zone in between the two galaxies highlighted as the
inner, shaded histogram. This shaded histogram is characterized by two
clear peaks in metallicity. However, if these clusters were all formed
approximately coevally (as implied by their derived ages) and in a
tightly confined spatial area, their metallicities {\it must} be very
similar; a clearly double-peaked distribution is therefore physically
unrealistic and therefore most likely another artefact caused by the
age--metallicity degeneracy. 

\begin{figure*}
\hspace*{0.2cm}
\psfig{figure=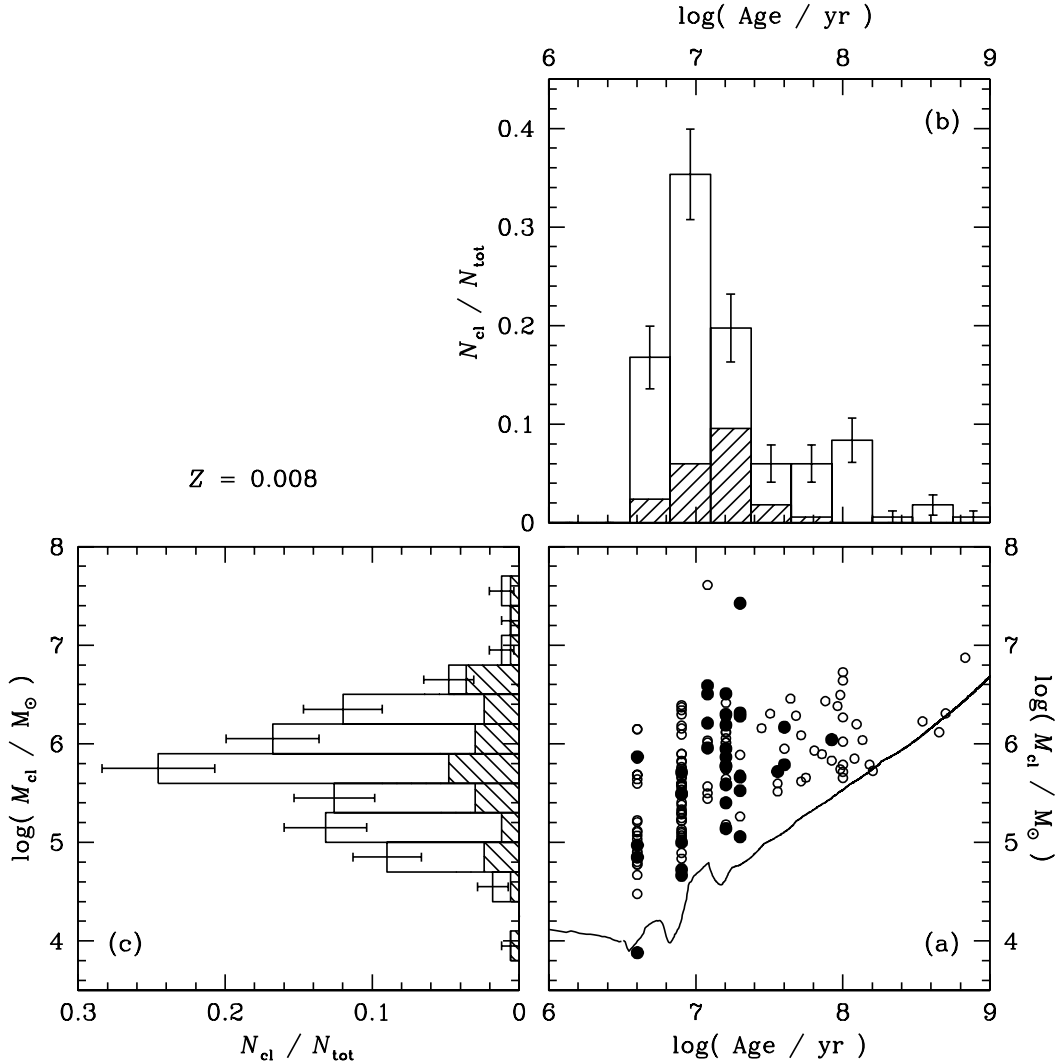,width=15cm}
\vspace*{-0.5cm}
\caption{\label{n6745fig1.fig}(a) -- Distribution of the NGC 6745
clusters in the (age vs. mass) plane, assuming a fixed metallicity of
$Z=0.008$ for all clusters. The solid bullets represent the clusters in
the most actively star forming region, i.e. NGC 6745b. Overplotted is
the expected, age-dependent detection limit, predicted by the Anders \&
Fritze--v. Alvensleben (2003) models for the appropriate stellar IMF. 
This model prediction is based on a detection limit of $m_{\rm F336W} =
23.5$ and $(m-M)_{{\rm NGC}6745} = 34.17$, assuming no extinction. (b)
and (c) -- Distributions of, respectively, the ages and masses of the
compact clusters in NGC 6745. The shaded histograms show the
distributions of the clusters in NGC 6745b. The uncertainties are of
the order of the histogram bin sizes.}
\end{figure*}

Based on these arguments, it is most likely that the NGC 6745 young
cluster system is best approximated by a metallicity consistent with
that of the low-metallicity peak in Fig.  \ref{n6745data.fig}c, at $Z
\lesssim 0.015$.  Therefore, we reapplied our SED fitting routines to
the observational data, assuming a more appropriate fixed metallicity of
$Z = 0.008$ for all clusters.  The resulting age and mass distributions,
are shown in Figs.  \ref{n6745fig1.fig}b and c.  We also show (in Fig. 
\ref{n6745fig1.fig}a) the clusters' locations in the (age vs.  mass)
plane with respect to our $\sim 90$\% completeness limit at $m_{\rm
F555W} \simeq 23.5$ mag (imposed by the shallow depth of the F336W
observations; cf.  Fig.  \ref{compl6745.fig}).  The new distribution of
extinction estimates, based on the $Z=0.008$ assumption, does not differ
significantly from that of Fig.  \ref{n6745data.fig}d. 

[We note that, while this procedure likely results in the most realistic
cluster parameters for the NGC 6745 system (due to the availability of
only four optical passbands), the situation for NGC 3310 is different. 
Because of the availability of observations of the NGC 3310 cluster
system covering a larger number of passbands and a greater wavelength
range, we concluded that fitting all of our free parameters
simultaneously resulted in the best parameter estimates.]

There is an enhancement of cluster formation around $\log( {\rm Age/yr}
) \simeq 7$, which is therefore our best estimate of the interaction
time-scale.  The clusters younger than $\log( {\rm Age/yr} ) \simeq 7.4$
($\sim 25$ Myr) are all located in either the interaction zone NGC 6745b
(see the shaded histogram in Fig.  \ref{n6745fig1.fig}b), or among the
sprinkling of blue objects resembling a spiral arm on the eastern edge
of the main galactic disc.  This is consistent with tidally-triggered
star (cluster) formation along the proposed trajectory of the companion
galaxy across the main disc component of NGC 6745 (see Sect. 
\ref{geometry.sect}).  The lower age cut-off at $\log( {\rm Age/yr} ) =
6.6$ (4 Myr) is artificial, and corresponds to the youngest ages
included in our SSP models (Schulz et al.  2002, Anders \& Fritze--v. 
Alvensleben 2003); a fraction of our NGC 6745 cluster sample may
therefore have even younger ages.  Thus, it seems likely that the
interaction began to have a significant effect on the star (cluster)
formation history some 25 Myr ago, and has been ongoing until the
present. 

We detected three compact clusters that are clearly associated with the
companion galaxy itself; they are unlikely to originate from the
interaction-induced starburst region, NGC 6745b, based on their large
projected distances from the interaction zone. Two of these have ages
of $\sim 100$ and $\sim 140$ Myr, and are located in the general field
of its disc. The third, located near the centre of the companion
galaxy, is of similar age as the starburst in NGC 6745a and b, $\sim 8$
Myr. Close inspection of the overall colour distribution of the field
star population in NGC 6745c reveals a bluer centre, which might
therefore imply that the tidal interaction with the main NGC 6745 galaxy
has also induced enhanced star formation in the companion galaxy, but at
a much lower level due to the absence of significant amounts of gas (see
Sect. \ref{geometry.sect}). The little available gas in NGC 6745c
might not have been sufficient for the formation of star {\it clusters}. 

There is no clear spatial dependence of the cluster masses, although the
lowest-mass clusters appear to originate in the lowest-density regions. 
This is most likely an observational selection effect, however. 
Although the galaxy's distance, combined with the photometric depth of
the observations, prevents us from probing below $\log( M_{\rm
cl}/M_\odot ) = 5.0$ for clusters older than $\sim 10$ Myr, we point out
that there is a significant population of high-mass clusters or cluster
complexes, with masses in the range $6.5 \lesssim \log( M_{\rm
cl}/M_\odot ) \lesssim 8.0$.  We have checked the reality and
signal-to-noise ratio of the SED of each of these objects individually,
and have only retained those objects that appear real beyond any
reasonable doubt.  These clusters do not have counterparts among the
Galactic GCs (e.g., Mandushev, Staneva \& Spassova 1991, Pryor \& Meylan
1993).  In fact, these masses are similar to or exceed the
spectroscopically confirmed mass estimates of the so-called ``super star
clusters'' (SSCs) in M82 (M82-F; Smith \& Gallagher 2001), and the
Antennae galaxies (Mengel et al.  2002).  Our detection of similarly
massive SSCs in NGC 6745, which are mostly located in the intense
interaction zone, NGC 6745b, supports the scenario that such objects
form preferentially in the extreme environments of interacting galaxies. 

We caution, however, that these massive SSC candidates may not be
gravitationally bound objects, but more diffuse star forming regions or
aggregates of multiple unresolved clusters instead.  Even with the
unprecedented {\sl HST} spatial resolution, we cannot distinguish
between these possibilities.  Nevertheless, we measure an effective
radius for the most massive object ($M_{\rm cl} \simeq 5.9 \times 10^8
M_\odot$) of only $R_{\rm eff} \sim 16$ pc.  However, this object
appears very elongated, or may in fact be a double cluster.  We should
keep in mind, of course, that this high mass estimate is a strong
function of the (low) metallicity assumed; if we had assumed solar
metallicity for this object, the derived age would have been
significantly smaller ($\sim 10-20$ Myr vs.  $\sim 1$ Gyr), and the mass
could be smaller by a factor of $\gtrsim 10$.  Even so, if we could
confirm this mass estimate spectroscopically, either of the
subcomponents would be the most massive cluster known to date,
significantly exceeding cluster W3 in NGC 7252, which has a mass of
about $(3-18) \times 10^7 M_\odot$, depending on the age, metallicity
and IMF assumed (Schweizer \& Seitzer 1998, Maraston et al.  2001). 

\subsection{The initial cluster mass distribution}
\label{n6745icmf.sect}

\begin{figure*}
\hspace*{1.2cm}
\psfig{figure=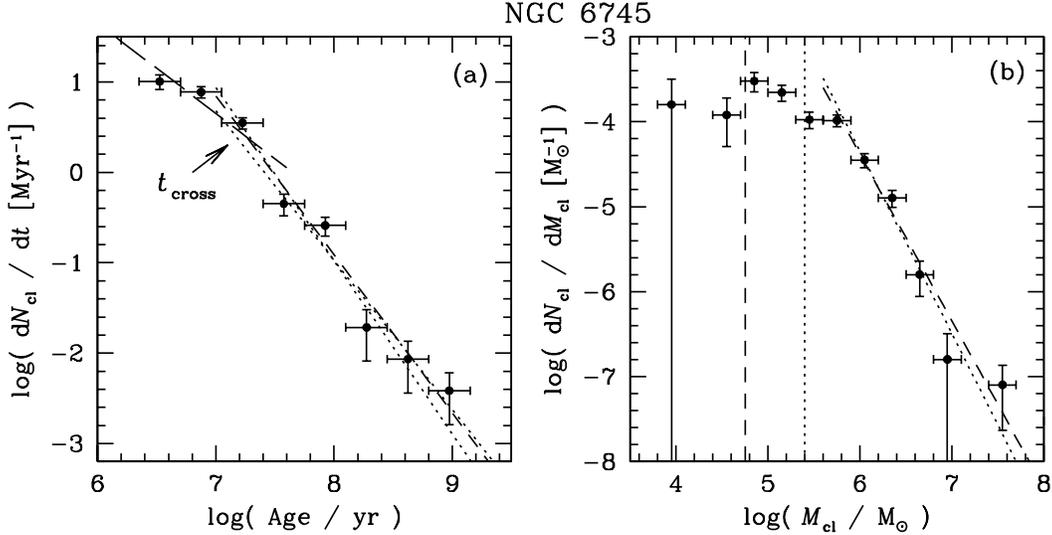,width=15cm}
\vspace*{-7cm}
\caption{\label{n6745disr.fig}(a) -- The cluster formation rate (in
number of clusters per Myr) in NGC 6745 as a function of age.  The
long-dashed line for $\log( {\rm Age / yr} ) \le 8.0$ is the expected
effect of a fading SSP, for a constant ongoing cluster formation rate,
shifted vertically to best match the data points for $\log( {\rm Age /
yr} ) \le 7.25$, and limited by the completeness of our F336W
observations.  The short-dashed and dotted lines are our best fit for
greater ages and the maximum likely deviation from this best fit,
respectively.  (b) -- Mass spectrum of the NGC 6745 clusters (number of
clusters per mass bin); line encoding as in panel {\it (a)}.  The
vertical dashed line at $\log( M_{\rm cl} / M_\odot ) = 4.75$
corresponds to our 90\% completeness limit for a
cluster age of $\log( {\rm Age / yr} ) = 7.0$.  The vertical dotted line
corresponds to the same completeness limit, but for a cluster age of
$\log( {\rm Age / yr} ) = 8.0$.}
\end{figure*}

As for NGC 3310, in Figs.  \ref{n6745disr.fig}a and b we show the NGC
6745 cluster formation rate (per linear time period) and its mass
spectrum, respectively.  For our analysis of the cluster disruption
time-scales in NGC 6745 we will make the same assumptions as described
in Sect.  \ref{ngc3310disr.sect}.  The long-dashed line in Fig. 
\ref{n6745disr.fig}a for the youngest ages, $\log( {\rm Age / yr} ) \le
8$, is the expected effect of a fading SSP in the F336W filter (which
determines our sample completeness for this galaxy), for a constant
ongoing cluster formation rate, shifted vertically to best match the
data points for $\log( {\rm Age / yr} ) \le 7.25$, and limited by the
completeness in the F336W observations.  For this band, the predicted
fading line is horizontal at $\log( {\rm Age/yr} ) \lesssim 6.5$ and has
a slope of $-1.12$ for older clusters, in the parameter space of Fig. 
\ref{n6745disr.fig}a.  The short-dashed lines in both Figs. 
\ref{n6745disr.fig}a and b are the best-fitting power-law slopes,
obtained for the age and mass ranges $\log( {\rm Age/yr} ) \ge 7.4$ and
$\log( M_{\rm cl}/M_\odot ) \ge 5.6$, while the dotted lines indicate
the likely uncertainties in the slopes.  The latter were obtained from
subsets of the data points used for the dashed fits, which were
determined over the full ranges.  The vertical dashed and dotted lines
in Fig.  \ref{n6745disr.fig}b at $\log( M_{\rm cl} / M_\odot ) = 4.75$
and $\log( M_{\rm cl} / M_\odot ) = 5.4$ correspond to our formal
completeness limit, for a peak young cluster age of $\log( {\rm Age /
yr} ) = 7.0$ and for the significant fraction of sample clusters with
ages up to $\log( {\rm Age / yr} ) = 8.0$, respectively.  Thus, it
follows that incompleteness plays at least some role for masses up to
$\log( M_{\rm cl}/M_\odot ) \sim 5.4$. 

For $\log( {\rm Age/yr} ) \ge 7.25$ we find a best-fitting slope of
$(1-\alpha)/\gamma = -1.86 (\pm 0.20)^{+0.21}_{-0.06}$, where the first
uncertainty represents the formal uncertainty in the fit and the second
the likely uncertainty resulting from our choice of fitting range (i.e.,
the dotted fits in Fig.  \ref{n6745disr.fig}a).  Assuming that $\alpha =
2.0$ (see below) we find that $\gamma = 0.54 \pm 0.06$, close to the
mean value of 0.62 derived by BL03 for cluster samples in four galaxies. 

The crossing point of the two power laws of the age distribution is at
$\log( t_{\rm cross} / {\rm yr} ) \simeq 7.3$.  Substituting these
values into Eq.  (15) of BL03, we obtain a characteristic disruption
time-scale for $10^4 M_\odot$ clusters of $\log( t_4^{\rm dis} / {\rm
yr} ) = 7.75$.  This implies that the disruption time-scale of clusters
more massive than $\sim 4 \times 10^5 M_{\odot}$ is longer than $5.5
\times 10^8$ yr.  All clusters in our sample, with one exception, are
younger than this.  The mass distribution of clusters more massive than
$4 \times 10^5 M_\odot$, shown in Fig.  \ref{n6745disr.fig}a, has a
slope of $-1.96(\pm0.15)\pm 0.19$, where the first uncertainty again
represents the formal uncertainty in the fit and the second the likely
uncertainty resulting from our choice of fitting range (i.e., the dotted
fits in Fig.  \ref{n6745disr.fig}b).  Since these massive clusters have
not yet suffered disruption, the measured slope represents that of the
cluster IMF of NGC 6745. 

\section{The Initial Cluster Mass Function}
\label{icmf.sect}

For star cluster systems as young as the tidally-induced cluster
populations in NGC 3310 and NGC 6745, for which the age of the starburst
in which they were formed is significantly smaller than their respective
characteristic cluster disruption time-scales, we have shown that the
application of the empirical cluster disruption models of BL03 results
in an independent estimate of the {\it initial} cluster mass function
slope, $\alpha$. For the NGC 3310 and NGC 6745 cluster systems, $\alpha
= 2.04(\pm 0.23)^{+0.13}_{-0.43}$ and $1.96 (\pm 0.15)\pm 0.19$,
respectively, for masses $M_{\rm cl} \gtrsim 10^5 M_\odot$ and $M_{\rm
cl} \gtrsim 4 \times 10^5 M_\odot$. 

The slopes of the CLFs of young star cluster systems, and to some lesser
extent those of their mass functions as well, have been studied
extensively -- in particular since the launch of the {\sl HST} -- in a
wide variety of star-forming environments, including interacting and
starburst galaxies, circumnuclear starburst rings, and also in
``normal'' spiral and irregular galaxies by means of the H{\sc ii}
region luminosity function (regarding the latter, see e.g. Kennicutt \&
Hodge 1980, Kennicutt et al. 1989, Arsenault et al. 1990, Cepa \&
Beckman 1990, Caldwell et al. 1991, Banfi et al. 1993, Elmegreen \&
Salzer 1999; and simulations by Oey \& Clarke 1998). In Table
\ref{slopes.tab}, we give an overview of the CLF (and, in a few cases,
of the mass function) slopes based on determinations available in the
literature, both for young cluster systems in interacting and starburst
galaxies, and for cluster populations in circumnuclear starburst rings. 

\begin{table*}
\caption[ ]{\label{slopes.tab}Cluster luminosity and mass function
slopes based on previous determinations in the literature
}
{\scriptsize
\begin{center}
\begin{tabular}{lcll}
\hline
\hline
\multicolumn{1}{c}{Galaxy} & \multicolumn{1}{c}{Slope ($-\alpha$)} &
\multicolumn{1}{c}{Notes} & \multicolumn{1}{c}{Reference}  \\
\hline
\multicolumn{4}{c}{A. Interacting and Starburst Galaxies} \\
NGC 1275 & $\sim -2$        & $M_V < -12$ & Holtzman et al. (1992) \\
         & $\sim -2$        & $M_B < -8$  & Carlson et al. (1998)  \\
NGC 1316 & $-1.7 \pm 0.1$   & 3 Gyr-old merger remnant & Goudfrooij et al. (2001) \\
         &                  &                          & Whitmore et al. (2002)   \\
NGC 3256 & $\sim -1.8$      & $M_B \lesssim -9.5$; $M_I \lesssim -10.5$ & Zepf et al. (1999) \\
NGC 3610 & $-1.78 \pm 0.05$ & $V < 26$; red, metal-rich clusters & Whitmore et al. (2002) \\
         & $-1.90 \pm 0.07$ & corrected for observational scatter \\
NGC 3921 & $-2.1 \pm 0.3$   & $M_V \le 8.5$ & Schweizer et al. (1996) \\
         & $-2.01 \pm 0.22$ & corrected for contamination by old GCs \\
NGC 4038/39&$-1.78 \pm 0.05$& $-15.5 \le M_V \le -10$; no completeness corrections & Whitmore \& Schweizer (1995) \\
         & $-2.12 \pm 0.04$ & $-14 < M_V < -8$, but contaminated by stars & Whitmore et al. (1999) \\
         & $-2.6  \pm 0.2$  & $-14 \le M_V \le -10.4$, corrected for contamination by stars \\
         & $-1.7  \pm 0.2$  & $-10.4 \le M_V \le -8$, corrected for contamination by stars \\
         & $-1.95 \pm 0.03$ & $6.4 < \log( {\rm Age/yr} ) < 6.8$; $10^4 \le M_{\rm cl} \le 10^6 M_\odot$ & Zhang \& Fall (1999) \\
         & $-2.00 \pm 0.08$ & $7.4 < \log( {\rm Age/yr} ) < 8.2$; $10^4 \le M_{\rm cl} \le 10^6 M_\odot$ \\
NGC 7252 & $\sim -2$        & $M_V \le -12$ & Whitmore et al. (1993) \\
         & $-1.90 \pm 0.04$ & $M_V \le -8$, inner sample & Miller et al. (1997) \\
         & $-1.80 \pm 0.07$ & $M_V \le -8$, outer sample \\
M51      & $-2.1 \pm 0.3$   & $2.5 \times 10^3 < M_{\rm cl} < 5 \times 10^4 M_\odot$; $\log( {\rm Age/yr} ) < 7$ & Bik et al. (2003) \\
         & $-2.0 \pm 0.05$  & $2.5 \times 10^3 < M_{\rm cl} < 2 \times 10^4 M_\odot$; $\log( {\rm Age/yr} ) < 7$ \\
M82 B    & $-1.2 \pm 0.3$   & $M_V \lesssim -11$ & de Grijs et al. (2001) \\
         & $-1.4 \pm 0.2$   & $8.6 < \log( {\rm Age/yr} ) < 9.1$ & Parmentier et al. (2003) \\
\\
\multicolumn{4}{c}{B. Circumnuclear Starburst Rings} \\
NGC 1097 & $\sim -2$        & $-14 \lesssim M_V \lesssim -11$ & Barth et al. (1995) \\
NGC 1512 & $\sim -2$        & $10^3 \lesssim M_{\rm cl} \lesssim 10^5 M_\odot$ & Maoz et al. (2001) \\
NGC 2997 & $\sim -2$        & & Maoz et al. (1996) \\
         & $-1.97 \pm 0.03$ & $M_V \le -9$  & Elmegreen et al. (1999) \\
NGC 3310 & $-2.2 \pm 0.03$  & $M_B < -10.5$ & Elmegreen et al. (2002) \\
         & $-2.4 \pm 0.04$  & from {\sl HST} {\it K}-band data \\
         & $-1.75 \pm 0.03$ & large-scale cluster complexes, {\it K} band \\
         & $-1.8 \pm 0.4$   & $17.7 \le m_{\rm F606W} \le 20.2$ & Paper I \\
NGC 5248 & $\sim -2$        & $10^3 \lesssim M_{\rm cl} \lesssim 10^5 M_\odot$ & Maoz et al. (2001) \\
NGC 6951 & $-2.12 \pm 0.04$ & $M_V \le -9$  & Elmegreen et al. (1999) \\
ESO 565-11&$-2.18 \pm 0.06$ & $M_V \le -8.9$; Cluster ages 4--6 Myr & Buta et al. (1999) \\
\hline
\end{tabular}
\end{center}
}
\end{table*}

In addition to these galaxies, Meurer et al. (1995) found that the CLF
slope of the combined sample of SSCs in 9 starburst galaxies, for
$M_{\rm F220W} \lesssim -14$, is consistent with $\alpha = 2$; Maoz et
al. (1996) constructed the combined CLF for the circumnuclear ring
clusters in NGC 1433, NGC 1512 and NGC 2997, based on UV observations,
and reached a similar conclusion for the high-luminosity end of their
CLF. Finally, Crocker et al. (1996) analysed the H{\sc ii} region
luminosity function in 32 (pseudo-)ringed S0 -- Sc galaxies, and found a
mean $\langle \alpha \rangle = 1.95 \pm 0.25$. 

Thus, in all but a few cases (e.g., M82 B), the CLF and, where derived,
the mass function slopes of the young clusters are similar ($\alpha
\simeq 2$) among a wide variety of galaxies, including those in which
the star and cluster formation process is dominated by strong tidal
forces, and systems of a more quiescent nature (e.g., normal spiral and
irregular galaxies; Kennicutt et al.  1989, Banfi et al.  1993,
Elmegreen \& Salzer 1999).  One should keep in mind that the observed
CLF is the result of both the original mass distribution of the young
star clusters and the giant molecular clouds (GMCs) they originated
from, and subsequent cluster disruption processes (e.g., Harris \&
Pudritz 1994, Bik et al.  2003, BL03, de Grijs et al.  2003b); in de
Grijs et al.  (2003b) and Parmentier et al.  (2003) we showed that the
M82 B cluster system must have undergone significant disruption owing to
the unusually short characteristic cluster disruption time-scale, $\log(
t_{\rm dis} / {\rm yr} ) \simeq 7.5 + 0.62 \times \log( M_{\rm cl}/10^4 M_\odot
)$ (de Grijs et al.  2003b), so that the observed shallow(er) CLF and
mass function slopes do not represent the {\it initial} cluster mass
function slope anymore.  However, in most other young cluster systems
known the expected characteristic cluster disruption time-scales are
well in excess of the median age of the cluster system, so that at least
the high-luminosity (high-mass) end of the CLF (and cluster mass
function) remains likely unaffected by significant alterations caused by
disruption (e.g., Buta et al.  1999, Zhang \& Fall 1999, Bik et al. 
2003, Paper I).  In this paper, we have derived very similar {\it mass}
function slopes, and have also properly accounted for the effects of
cluster disruption. 

Under the usual assumption that the stellar mass-to-light ratio does not
vary significantly over the age range of a given young cluster system,
or -- alternatively -- after correcting the observed CLF to a common age
(e.g., Meurer 1995, Fritze--v. Alvensleben 1999, de Grijs et al. 2001,
2003a,b), the resulting CLF slope can be interpreted in terms of the
intrinsic cluster mass distribution. The observed mass function slope,
$\alpha \simeq 2$, is very similar to those of the mass functions of
Galactic open star clusters (e.g., van den Bergh \& Lafontaine 1984,
Elson \& Fall 1985), GMCs (e.g., Casoli, Combes \& Gerin 1984, Sanders,
Scoville \& Solomon 1985, Solomon et al. 1987, Solomon \& Rivolo 1989,
Williams \& McKee 1997) and GMC cores (e.g., Harris \& Pudritz 1994,
Williams, de Geus \& Blitz 1994, McLaughlin \& Pudritz 1996 and
references therein; see also the review by Blitz 1991), and to that of
the young compact cluster system in the Large Magellanic Cloud (LMC;
e.g., Elson \& Fall 1985, Elmegreen \& Efremov 1997). Thus, it appears
that the formation of young compact star clusters, such as the ones
discussed in the current series of papers, and of open clusters, GMCs
and stellar OB associations (e.g., Elmegreen \& Efremov 1997, Elmegreen
2002) is driven by similar physical processes, which results naturally
from models advocating that the mass distribution of the progenitor GMCs
is well approximated by a turbulence-driven fractal structure, allowing
cloud growth by agglomeration (e.g., Harris \& Pudritz 1994, Elmegreen
\& Efremov 1997, Elmegreen et al. 1999, Elmegreen 2002). 

\section{Summary and Conclusions}
\label{conclusions.sect}

Both NGC 3310 and NGC 6745 are examples of nearby interacting galaxies
undergoing intense, tidally-triggered starbursts. The production of
luminous, compact star clusters seems to be a hallmark of intense star
formation. It is likely that a large fraction of the star formation in
starbursts takes place in the form of such concentrated clusters. The
basic cluster properties, including their sizes, luminosities, and -- in
several cases -- spectroscopic mass estimates are entirely consistent
with what is expected for young Galaxy-type GCs. Young compact star
clusters are therefore important because of what they can tell us about
GC formation and evolution. They are also important as probes of the
history of star formation, chemical evolution, IMF, and other physical
characteristics in starbursts. 

In this paper we have used the basic properties of the rich young star
cluster systems in both starburst galaxies, including estimates of the
individual cluster ages, masses, metallicities, and extinction values,
to derive the cluster formation history and the subsequent
evolution of the star cluster systems as modified by the effects of
cluster disruption. Cluster disruption processes must be taken into
account for the determination of the cluster formation history from the
age distribution of a magnitude-limited cluster sample, because the
observed age distribution is that of the surviving clusters only. 

For the NGC 3310 star cluster system we used the basic properties
obtained in Paper I, which were based on the detailed comparison of
updated model SSPs (including the contributions of gaseous, nebular
emission) with the clusters' SEDs derived from broad-band {\sl HST}
imaging observations.  We applied our improved understanding of the
systematic uncertainties involved in such an exercise (Paper I) to
simultaneously determine robust cluster ages, masses, metallicities and
extinction values for the tidally-induced young star cluster system in
NGC 6745, based on imaging observations in the minimum of four
broad-band passbands covering the entire optical wavelength range from
the {\sl HST}-equivalent {\it U} to {\it I}-band filters.  We further
expanded our analysis of the systematic uncertainties involved in this
type of analysis by examining the effects of {\it a priori} assumptions
of the individual cluster metallicities.  We find that -- if
observations covering a sufficiently large optical/NIR wavelength range
are available -- (i) it is very important to determine all of the free
parameters (age, metallicity, and extinction, and the corresponding mass
estimates) {\it independently} for each individual cluster, instead of
assuming a generic value for any of these parameters (such as assuming a
fixed metallicity); and (ii) the widths of the resulting age and mass
distributions, under the assumption of a fixed metallicity for all
clusters, are broader than their intrinsic widths owing to the
propagation of model and measurement uncertainties. 

The central morphology of NGC 3310 is dominated by a bright starburst
ring containing a large number of young star clusters. The age (and
metallicity) distributions of both the clusters in the ring and of the
sample of clusters outside the ring are statistically indistinguishable,
with a peak in the age distribution at $\log( {\rm Age/yr} ) \simeq
7.45$, but the ring clusters appear to peak at slightly greater masses
than those found outside the ring. The overabundance of low-mass
clusters outside the ring compared to their counterparts in the
starburst ring might (at least partially) be due to a spatial dependence
of the completeness fraction, but there is a clear and significant
excess of higher-mass clusters {\it in} the ring compared to the
non-ring cluster sample, even after taking the systematic uncertainties
in our mass estimates into account. It appears that the physical
conditions in the starburst ring, such as caused by the effects of the
proposed bar-driven instabilities, the higher density of the ISM, and
the associated higher likelihood to encounter significant propagating
shock waves, may be conducive for the formation of higher-mass star
clusters, on average, than in the relatively more quiescent environment
of the main galactic disc. 

We furthermore conclude that the starburst producing the NGC 3310 ring
clusters has been ongoing for at least the past 40 Myr, at an
approximately constant cluster formation rate during the burst period. 
Star cluster formation has proceeded at a similar rate in the central
disc of the galaxy; due to the extreme youth of both samples of
starburst-induced star clusters, statistically significant differences
between the age distributions of the ring and non-ring populations have
not yet had time to develop, which implies that we are still observing
at least part of the cluster sample in the environment determined by
their initial conditions. 

For the NGC 6745 cluster system we derive an median age of $\sim 10$
Myr, assuming a fixed metallicity of $Z = 0.008$ to avoid the
significant effects of the age--metallicity degeneracy for
these young ages. Based on the age distribution of the star clusters,
and the H{\sc i} morphology of the interacting system as observed with
the VLA, we qualitatively confirm the interaction-induced enhanced star
formation scenario. NGC 6745 contains a significant population of
high-mass clusters, with masses in the range $6.5 \lesssim \log( M_{\rm
cl}/M_\odot ) \lesssim 8.0$. These clusters do not have counterparts
among the Galactic GCs, but are similar to or exceed the
spectroscopically confirmed mass estimates of the SSCs in M82 and the
Antennae galaxies. Our detection of such massive SSCs in NGC 6745,
which are mostly located in the intense interaction zone, NGC 6745b,
supports the scenario that such objects form preferentially in the
extreme environments of interacting galaxies. 

For star cluster systems as young as the tidally-induced cluster
populations in NGC 3310 and NGC 6745, for which the age of the starburst
in which they were formed is significantly lower than their respective
characteristic cluster disruption time-scales (of, respectively,
$\log(t_4^{\rm dis}/{\rm yr}) = 8.05$ and 7.75), we have shown that the
application of the empirical cluster disruption models of BL03 results
in an independent estimate of the {\it initial} cluster mass function
slope, $\alpha$. For the NGC 3310 and NGC 6745 cluster systems, $\alpha
= 2.04(\pm 0.23)^{+0.13}_{-0.43}$ and $1.96 (\pm 0.15)\pm 0.19$,
respectively, for masses $M_{\rm cl} \gtrsim 10^5 M_\odot$ and $M_{\rm
cl} \gtrsim 4 \times 10^5 M_\odot$. These mass function slopes are
consistent with those of other young star cluster systems in interacting
and starburst galaxies, circumnuclear starburst rings, and of the H{\sc
ii} region luminosity (and mass) functions in ``normal'' spiral and
irregular galaxies. They are also very similar to the mass function
slopes of Galactic open clusters and (OB) associations, of GMCs and
their cores, and of the young compact clusters in the LMC. It has been
shown previously that this can be understood if the mass function of the
progenitor GMC is approximated by a turbulence-driven fractal structure,
allowing cloud growth by agglomeration. 

\section*{acknowledgments} This paper is based on archival observations
with the NASA/ESA {\sl Hubble Space Telescope}, obtained at the Space
Telescope Science Institute (STScI), which is operated by the
Association of Universities for Research in Astronomy (AURA), Inc.,
under NASA contract NAS 5-26555. This paper is also partially based on
ASTROVIRTEL research support, a project funded by the European
Commission under 5FP Contract HPRI-CT-1999-00081. This research has
made use of NASA's Astrophysics Data System Abstract Service. RdG
acknowledges partial support from The British Council under the {\sl
UK--Netherlands Partnership Programme in Science}, additional funding
from the Particle Physics and Astronomy Research Council (PPARC), and
hospitality at the Astronomical Institute of Utrecht University on
several visits. PA is partially supported by DFG grant Fr 916/11-1; PA
also acknowledges partial funding from the Marie Curie Fellowship
programme EARASTARGAL ``The Evolution of Stars and Galaxies'', funded by
the European Commission under 5FP contract HPMT-CT-2000-00132.

\end{document}